\documentclass[11pt]{article}
\usepackage{graphicx}
\usepackage{longtable,lscape}
\usepackage{amsfonts}
\usepackage{float}
\usepackage{url}
\newcommand{\compl}{{\mathbb C}}
\newcommand{\real}{{\mathbb R}}
\newcommand{\captionfonts}{\footnotesize}
\makeatletter  
\long\def\@makecaption#1#2{%
  \vskip\abovecaptionskip
  \sbox\@tempboxa{{\captionfonts #1: #2}}%
  \ifdim \wd\@tempboxa >\hsize
    {\captionfonts #1: #2\par}
  \else
    \hbox to\hsize{\hfil\box\@tempboxa\hfil}%
  \fi
  \vskip\belowcaptionskip}
\makeatother   
\begin{document}
\title{Quantum Particles as Conceptual Entities: A Possible Explanatory Framework for Quantum Theory}
\author{Diederik Aerts\\
        \normalsize\itshape
        Center Leo Apostel for Interdisciplinary Studies \\
        \normalsize\itshape
        and Departments of Mathematics and Psychology \\
        \normalsize\itshape
        Vrije Universiteit Brussel, 1160 Brussels, 
       Belgium \\
        \normalsize
        E-Mail: \textsf{diraerts@vub.ac.be}
        }
\date{}
\maketitle              
\begin{abstract}
\noindent We put forward a possible new interpretation and explanatory framework for quantum theory. The basic hypothesis underlying this new framework is that quantum particles are conceptual entities. More concretely, we propose that quantum particles interact with ordinary matter, nuclei, atoms, molecules, macroscopic material entities, measuring apparatuses, \ldots, in a similar way to how human concepts interact with memory structures, human minds or artificial memories. We analyze the most characteristic aspects of quantum theory, i.e. entanglement and non-locality, interference and superposition, identity and individuality in the light of this new interpretation, and we put forward a specific explanation and understanding of these aspects. The basic hypothesis of our framework gives rise in a natural way to a Heisenberg uncertainty principle which introduces an understanding of the general situation of `the one and the many' in quantum physics. A specific view on macro and micro different from the common one follows from the basic hypothesis and leads to an analysis of Schr\"odinger's Cat paradox and the measurement problem different from the existing ones. We reflect about the influence of this new quantum interpretation and explanatory framework on the global nature and evolutionary aspects of the world and human worldviews, and point out potential explanations for specific situations, such as the generation problem in particle physics, the confinement of quarks and the existence of dark matter.
\end{abstract}

\section{Introduction}
We have formulated a proposal for a possible new interpretation of quantum theory accompanied by a specific explanatory framework \cite{aerts2009a}. In the present article we elaborate this new interpretation and its explanatory framework, as well as its consequences for the micro and macroscopic world.

The basis of our new quantum interpretation is the hypothesis that a quantum particle is a conceptual entity, more specifically, that a quantum particle interacts with ordinary matter in a similar way than a human concept interacts with a memory structure. With ordinary matter we mean substance made of elementary fermions, i.e. quarks, electrons and neutrinos, hence including all nuclei, atoms, molecules, macroscopic material entities and hence also measuring apparatuses. Ordinary matter is sometimes also called baryonic matter in the literature, when contrasted with dark matter, which plausibly is not constituted of baryons. A memory structure for human concepts can be a human mind or an artificial memory.

The idea for this basic hypothesis follows from our work involving the use of quantum formalism for the modeling of human concepts \cite{aerts2009b,aerts2009c,aertsaertsgabora2009,aertsdhooghe2009,aerts2007a,aerts2007b,aertsczachordhooghe2006,aertsgabora2005a,aertsgabora2005b,aertsczachor2004,gaboraaerts2002,aertsaerts1994}. This made us ask the question that, `if quantum mechanics as a formalism models human concepts so well, perhaps this indicates that quantum particles themselves are conceptual entities?' More importantly, however, the specific way in which quantum mechanics models human concepts and the fact that this yields a simple explanation for both interference and entanglement, shedding completely new light on the underlying issues of identity, indistinguishability and individuality with respect to quantum particles, have given us sufficient reason to seriously consider the possibility of using these findings to propose a possible new interpretation and explanation of quantum theory. 

The spirit in which we put forward this possible new interpretation and explanatory framework for quantum theory is one of explicit humbleness. Indeed, we are well aware that the suggestion of `a possible new interpretation and explanatory framework for quantum theory' carries a big load and responsibility. Existing interpretations of quantum theory have been scrutinized for many decades, theoretically as well as experimentally, and introducing a new quantum interpretation should not be undertaken light-heartedly. That is why our decision to write down the material in this article and in \cite{aerts2009a} was preceded by a long spell of hesitation. What eventually tipped the scales was the thought that making the idea and the results available to the scientific community for reflection and comments would subject it to the scrutiny it needed. And we should add one more thing. If we look at all currently existing interpretations of quantum physics, we can easily see that it defies any kind of explicatory framework. We need but recall Richard Feynman's famous verdict: `I think I can safely say that nobody understands quantum mechanics \cite{feynman1967}', or his more specific view that `things on a very small scale [like electrons] behave like nothing that you have any direct experience about. They do not behave like waves, they do not behave like particles, they do not behave like clouds, or billiard balls, or weights on springs, or like anything that you have ever seen \cite{feynmanleightonsands1963}'. Starting from the hypothesis referred to above, the quantum interpretation we propose offers a very clear and explicit explanation, namely that `quantum particles behave like concepts'. This means that, if proven correct, this new quantum interpretation would provide an explanation according to which `quantum particles behave like something we are all very familiar with and have direct experience with, namely concepts'. The explanatory framework resulting from this possible new interpretation could thus lead to a fundamentally new understanding of quantum theory. In \cite{aerts2009a}, we analyzed some of the major phenomena in quantum theory traditionally classified as `not understood', such as interference, entanglement, the issues of identity, indistinguishability and individuality and Schr\"odinger's Cat paradox and/or the measurement problem, showing how this new interpretation may help to understand their nature. In the present article, we will analyze these principal aspects of quantum theory more in detail, while looking into the impact of the suggested explanatory framework on other aspects of quantum theory as well as on the nature of our global worldview. 

\section{Concept Combination and Quantum Entanglement} \label{entanglement}
In this section we analyze how entanglement and non-locality can be understood starting from the basic hypothesis that quantum particles behave like concepts. We show that the way concepts combine naturally gives rise to the presence of entanglement and non-locality, mathematically described in a way similar to how entanglement and non-locality are described for quantum entities. This becomes evident if we have a close look at situations in which the wave picture for quantum entities fails to adequately model quantum entanglement and non-locality.

\subsection{Neither Particles Nor Waves}
Waves and particles have historically played a major role in attempts to understand the behavior of quantum entities. The reason is that two types of observations occur with respect to quantum entities in experimental situations, viz. clicks of detectors and spots on detection screens, indicating particle aspects of the entities that are being considered, and interference and diffraction patterns, indicating wave aspects of these entities.

However, the archetypal use of particles and waves to explain experimental phenomena in physics dates back much further than quantum mechanics. More specifically, with respect to understanding the nature of light, there has been a real competition between particle and wave models \cite{achinstein1991}. The earliest theory of light in the seventieth century, supported by Ren\'e Descartes, Robert Hooke and others, and worked out in most detail by Christian Huygens, was a wave theory \cite{huygens1900}. This wave theory of light was soon to be overshadowed by Isaac Newton's particle theory of light \cite{newton1979}. Waves showed up again as an explanatory model for the behavior of light when Thomas Young introduced the thought experiment later to become known as the double-slit experiments, mentioning for the first time the idea of interference \cite{young1802}. However, Young's arguments did not yet change the tide in favor of waves. This happened later, in the early nineteenth century, as a consequence of the work of Augustin Fresnel, who elaborated a mathematical description of diffraction \cite{fresnel1819}. The wave theory of light became fully accepted in the course of the nineteenth century, and James Clerk Maxwell elaborated a model in which light appeared  as electromagnetic waves whose behavior was governed by a set of equations, now called Maxwell's equations.
In 1901, Max Planck put forward the idea of quantized energy of the oscillators describing the electromagnetic radiation to solve a serious problem with the observed law of radiation of a heated black body \cite{planck1901}. In 1905, Albert Einstein proposed a description of the photoelectric effect, unexplainable by the wave theory of light, by using Planck's hypothesis and also straightforwardly postulating the existence of photons, i.e. quanta of light energy with particular properties \cite{einstein1905}. In 1924, Louis de Broglie introduced the hypothesis that all matter, not just light, entails a wave structure, and three years later de Broglie's hypothesis was confirmed for electrons in a diffraction experiment \cite{debroglie1923}. The modern quantum theory was formulated by Werner Heisenberg in the discrete and particle-like setting of matrix mechanics \cite{heisenberg1925} and a year later by Erwin Schr\"odinger in the continuous and wave-like setting of wave mechanics \cite{schrodinger1926a}. Both where proven to be equivalent as abstract models \cite{schrodinger1926b}, which made it possible for John von Neumann later to formulate their abstract version as Hilbert space quantum mechanics \cite{vonneumann1932}. The Copenhagen interpretation of quantum theory was worked out by Heisenberg and Bohr in the years following 1925, and one of the basic ideas is that of particle-wave duality, namely that quantum entities entail wave or particle properties depending on the type of experiment to be performed \cite{heisenberg1927,bohr1928}. Louis de Broglie and later David Bohm elaborated a pilot wave construct to account for the observed particle-wave duality. In this view, a particle {\it and} a wave are connected to a quantum entity, the particle has a well-defined position and momentum, and it is guided by a wave function derived from Schr\"odinger's equation \cite{debroglie1928,bohm1952}.

Although there have always been signs that quantum entities are perhaps neither particles nor waves, there has not been any explicit evidence as to their true nature. Let us concentrate on what we think is the greatest obstacle to the `wave view' of quantum entities. If one considers two quantum entities $S_1$ and $S_2$, described by wave functions $\psi_1(x_1,y_1,z_1)$ and $\psi_2(x_2,y_2,z_2)$,  respectively, which are complex functions of three real variables, then the joint quantum entity consisting of both entities is described by a wave function $\psi_{1,2}(x_1,y_1,z_1,x_2,y_2,z_2)$, which is a complex function of six real variables. This function $\psi_{1,2}(x_1,y_1,z_1,x_2,y_2,z_2)$ is in general not the product of two complex functions of three real variables. In the abstract mathematical formulation of von Neumann \cite{vonneumann1932}, the state $\psi_1(x_1,y_1,z_1)$ of the first quantum entity $S_1$ is an element of the Hilbert space $L^2(\real^3)_1$ of all square integrable complex functions of three real variables, and the state of the second quantum entity $S_2$ is an element of the same Hilbert space $L^2(\real^3)_2$ of all square integrable complex functions of three real variables. The state $\psi_{1,2}(x_1,y_1,z_1,x_2,y_2,z_2)$ of the joint quantum entity consisting of $S_1$ and $S_2$ is an element of the Hilbert space $L^2(\real^6)$ of all square integrable complex functions of six real variables. From the mathematics of Hilbert spaces it follows that $L^2(\real^6)$ is isomorphic to $L^2(\real^3)\otimes L^2(\real^3)$, where $\otimes$ stands for `tensor product'. This is the reason why in abstract formulations of quantum theory the tensor product of Hilbert spaces describes joint quantum entities. 

Many experiments have by now confirmed in great detail the correctness of this quantum procedure to describe joint quantum entities by means of the tensor product, so that there is no doubt about its validity. Moreover, it is exactly the situations where the wave function of the joint entity of two quantum entities is not a product of wave functions of the two constituent entities that give rise to the quantum phenomenon of entanglement and non-locality. More specifically, we can detect Einstein Podolsky Rose type of correlations that violate Bell's inequalities \cite{einsteinpodolskyrosen1935,bell1964,aspectgrangierrobert1981,tittelbrendelgisinherzogzbindengisin1998,weihsjenneweinsimonweinfurterzeilinger1998}. In view of all this, quantum mechanics of more than one quantum entity brings to evidence that the wave function does not correspond to a wave in three-dimensional space, so that it may not correspond to a wave at all. Of course, we are saying nothing new here, many have pointed out the difficulty encountered with the wave view for the situation of many quantum entities, but even so the wave view has remained one of the basic ingredients of many interpretations. This author remembers a personal conversation with John Bell in the eighties of the previous century, where Bell's response to this difficulty was that he preferred considering the possibility of coping with real waves but in more than three dimensions to having to give up any picture altogether, which explains Bell's support of the de Broglie Bohm interpretation in these years.

\subsection{Violating Bell's Inequalities} \label{violationbellsinequalities}
We want to show now that if quantum entities are considered to be concepts rather than objects, and hence neither particles nor waves, the type of structure provoking entanglement and non-locality appears in a natural way. We do this by considering a concrete example of how concepts that are combined give rise to entanglement and non-locality, and we will also use this example to explain in further detail our new interpretation and explanatory framework for quantum theory.

Before we proceed we want to explain some basic aspects of the quantum modeling scheme for concepts which we worked out in \cite{aertsgabora2005a,aertsgabora2005b,gaboraaerts2002}. One of its fundamental aspects is the introduction of the notion of `state of a concept'. Consider the concept {\it Fruits}. We introduce the notion of `state', such that the state of the concept {\it Fruits} is identified experimentally by measuring the typicality weights of exemplars and the application values of features of {\it Fruits}. Such measurements are standard practice in psychological concept research and detailed examples with references to the psychology literature are given in \cite{aertsgabora2005a,aertsgabora2005b}. For our analysis of interference in section \ref{interferencesuperposition}, we have explicitly  used data of typicality weights for {\it Fruits} measured in \cite{hampton1988}. More specifically, the second column of Table 1 represent the typicality weights measured in \cite{hampton1988} of the different exemplars of {\it Fruits} shown in the first column of Table 1. {\it Apple} was elected to be `the most typical fruit' of the considered group of exemplars, with typicality weight 0.1184, followed by {\it Elderberry} with typicality weight 0.1138. The least typical fruit from the list was chosen to be {\it Lentils}, with typicality weight 0.0095 and the second least was {\it Garlic}, with typicality weight 0.0100. A change of the state of {\it Fruits} would be provoked by having another concept, for example the concept {\it Tropical}, combine with it to give {\it Tropical Fruits}. The exemplar {\it Coconut} would raise its typicality weight for this state of {\it Fruits} and most probably score higher than {\it Apple} and {\it Elderberry}. To know the effect on the rest from the list of exemplars considered in Table 1, the experiment should be performed, but analogous experiments performed in \cite{aertsgabora2005a} show that the effect is considerable. The concept {\it Tropical} plays the role of a context in the combination {\it Tropical Fruits}, and as a context changes the state of {\it Fruits} to that of {\it Tropical Fruits}. It is this type of change, and the probability connected to it, that we have modeled by using the quantum formalism \cite{aertsgabora2005a,aertsgabora2005b,gaboraaerts2002}. Of course, many examples of `change of state' of the concept {\it Fruits} can be given. {\it Juicy} can act as a context, and combined with {\it Fruits}, yielding {\it Juicy Fruits}, it will give rise to another field of typicality values with respect to the set of exemplars considered in Table 1. But also more elaborate and more complex contexts can be considered. For example in the combination of concepts {\it This Fruits is too Old to be Eaten}, the combination of concepts {\it Too Old to be Eaten} plays the role of context, changing the state of {\it Fruits}. Or again, in the combination of concepts  {\it Look there, he Mistakingly Thinks that he is Eating a Fruit}, the context {\it Mistakingly Thinks he is Eating} changes the state of {\it Fruits} quite dramatically. We would not be surprised that tests of exemplars like {\it Garlic}, whose typicality for the default state of {\it Fruits} is very low, would yield a high typicality score. In \cite{aertsgabora2005a} we explicitly tested these drastic changes of state. We should add that also exemplars themselves are states of the concept of which they are an exemplar. We will discuss the subtleties of these matters in greater detail in sections \ref{abstractconcrete} and \ref{oneandmany}. 

Let us now introduce an example that violates Bell's inequalities. Consider the two concepts {\it Animal} and {\it Food} and how they can be combined into a conceptual combination {\it The Animal eats the Food}. Our aim is to formulate correlation experiments violating Bell's inequalities and analyze in which way entanglement and non-locality appear when concepts are combined. We also want to show how this inspires the explanatory framework for quantum mechanics that we put forward.
To formulate the experiment that violates Bell's inequalities, we consider the following exemplars of these two concepts. For the concept {\it Animal} we consider the couples of exemplars,  {\it Cat}, {\it Cow} and {\it Horse}, {\it Squirrel}, and for the concept {\it Food}, the couples of exemplars {\it Grass}, {\it Meat} and {\it Fish}, {\it Nuts}. Our first experiment $A$ consists in test subjects choosing for the concept {\it Animal} between one of the two exemplars {\it Cat} or {\it Cow}, and responding to the question `which is a good example of {\it Animal}?'. We put $E(A)=+1$ if {\it Cat} is chosen, and $E(A)=-1$ if {\it Cow} is chosen, introducing in this way the function $E$ which measures the `expectation value' for the test outcomes concerned. Our second experiment $A'$ consists in test subjects choosing for the concept {\it Animal} between one of the two exemplars {\it Horse} or {\it Squirrel}, and responding to the same question. We consistently put $E(A')=+1$ if {\it Horse} is chosen and $E(A')=-1$ if {\it Squirrel} is chosen to introduce a measure of the expectation value. The third experiment $B$ consists in test subjects choosing for the concept {\it Food} between one of the two exemplars {\it Grass} or {\it Meat}, and responding to the question `which is a good example of {\it Food}?'. We put $E(B)=+1$ if {\it Grass} is chosen and $E(B)=-1$ if {\it Meat} is chosen, and the fourth experiment $B'$ consists in test subjects choosing for the concept {\it Food} between one of the two exemplars {\it Fish} or {\it Nuts}, and responding to the same question. We put $E(B')=+1$ if {\it Fish} is chosen and $E(B')=-1$ if {\it Nuts} is chosen.

Now we consider coincidence experiments in combinations $AB$, $A'B$, $AB'$ and $A'B'$ for the conceptual combination {\it The Animal eats the Food}. Concretely, this means that, for example, test subjects taking part in the experiment $AB$, choose between the four possibilities (1) {\it The Cat eats the Grass}, (2) {\it The Cow eats the Meat}, and if one of these is chosen we put $E(AB)=+1$, or (3) {\it The Cat eats the Meat}, (4) {\it The Cow eats the Grass}, and if one of these is chosen we put $E(AB)=-1$, responding to the question `which is a good example of {\it The Animal eats the Food}'. For the coincidence experiment $A'B$ subjects choose between (1) {\it The Horse eats the Grass}, (2) {\it The Squirrel eats the Meat}, and in case one of these is chosen we put $E(A'B)=+1$, or (3) {\it The Horse eats the Meat}, (4) {\it The Squirrel eats the Grass}, and in case one of these is chosen we put $E(A'B)=-1$, responding to the same question `which is a good example of {\it The Animal eats the Food}'. For the coincidence experiment $AB'$ subjects choose between (1) {\it The Cat eats the Fish}, (2) {\it The Cow eats the Nuts}, and in case one of these is chosen we put $E(AB')=+1$, or (3) {\it The Cow eats the Fish}, (4) {\it The Cat eats the Nuts}, and in case one of these is chosen we put $E(AB')=-1$, responding to the same question. And finally, for the coincidence experiment $A'B'$ subjects choose between (1) {\it The Horse eats the Fish}, (2) {\it The Squirrel eats the Nuts}, and in case one of these is chosen we put $E(A'B')=+1$, or (3) {\it The Horse eats the Nuts}, (4) {\it The Squirrel eats the Fish}, and in case one of these is chosen we put $E(A'B')=-1$, responding to the same question.

Quite obviously, in coincidence experiment $AB$, both {\it The Cat eats the Meat} and {\it The Cow eats the Grass} will yield rather high scores, with the two remaining possibilities {\it The Cat eats the Grass} and {\it The Cow eats the Meat} being chosen less. This means that we will get $E(AB)\approx-1$. On the other hand, in the coincidence experiment $A'B$ one of the four choices will be prominent, namely {\it The Horse eats the Grass}, while the three other possibilities {\it The Squirrel eats the Meat}, {\it The Horse eats the Meat}, and {\it The Squirrel eats the Grass} will be chosen much less frequently by the test subjects. This means that we have $E(A'B)\approx+1$. In the two remaining coincidence experiments, we equally have that only one of the choices is prominent. For $AB'$, this is {\it The Cat eats the Fish}, with the other three {\it The Cow eats the Nuts}, {\it The Cow eats the Fish} and {\it The Cat eats the Nuts} being chosen much less. For $A'B'$, the prominent choice is {\it The Squirrel eats the Nuts}, while the other three {\it The Horse eats the Fish}, {\it The Horse eats the Nuts} and {\it The Squirrel eats the Fish} are chosen much less often. This means that we have $E(AB')\approx+1$ and $E(A'B')\approx+1$. If we now substitute the different expectation values related to the coincidence experiments in the Clauser-Horne-Shimony-Holt variant of Bell's inequalities \cite{clauserhorneshimonyholt1969}, we get
\begin{equation}
E(A'B')+E(A'B)+E(AB')-E(AB)\approx+4
\end{equation}
Since the Clauser-Horne-Shimony-Holt variant of Bell's inequalities requires this expression to be contained in the interval $[-2,+2]$, our calculation shows that our example constitutes a strong violation of Bell's inequalities.

We do not doubt that an experiment involving test subjects will yield data that violate Bell's inequalities. However, rather than through experiments, we have opted for collecting relevant data using the World Wide Web. The reason for this is that it sheds light on an interesting aspect of our new quantum interpretation and our explanatory framework, as we will show further on in this article. We will now first explain how we have set about this and then we analyze the data collected.

Using Google we establish the numbers of pages that contain the different combinations of items as they appear in our example. More concretely, for the coincidence experiment $AB$, we find $1,550$ pages that contain the sentence part `cat eats grass', $125$ pages that contain `cow eats meat', $457$ pages that contain `cat eats meat' and $4,240$ pages that contain `cow eats grass'. This means that on a totality of $1,550+125+457+4,240=6,372$ pages, we get the fractions of $1,550$, $125$, $457$ and $4,240$ for the various combinations considered. We suppose that each page is elected with equal probability, which allows us to calculate the probability for one of the combinations to be elected. This gives $P(A_1,B_1)=1,550/6,372=0.2433$ for `cat eats grass', $P(A_2,B_2)=125/6,372=0.0196$ for `cow eats meat', $P(A_1,B_2)=457/6,372=0.0717$ for `cat eats meat' and $P(A_2,B_1)=4,240/6,372=0.6654$ for `cow eats grass'. We take for granted that the number of pages containing two of such expressions is negligible -- although Google has no way of verifying this --, which means that we do not have to take this into account for the calculation of the probabilities. Anyhow, even if this was not so, and a more complicated manner to calculate the probabilities was needed, this would not affect the outcome, i.e. the violation of Bell's inequalities, as can be inferred from the following of our analysis. Knowing these probabilities, we can again calculate the expectation value for this coincidence experiment by means of the equation $E(A,B)=P(A_1,B_1)+P(A_2,B_2)-P(A_2,B_1)-P(A_1,B_2)=-0.4743$. We calculate the expectation values $E(A',B)$, $E(A,B')$ and $E(A',B')$ in an analogous way making use of the Google results. For the coincidence experiment $A'B$ we find $768$ pages that contain `horse eats grass', $36$ pages with `squirrel eats meat', $6$ pages with `horse eats meat' and $0$ pages with `squirrel eats grass'. This gives $P(A'_1,B_1)=0.9481$, $P(A'_2,B_2)=0.0444$, $P(A'_1, B_2)=0.0074$ and $P(A'_2,B_1)=0$ and $E(A',B)=0.9852$. For the coincidence experiment $AB'$ we find $1,040$ pages that contain `cat eats fish', $2$ pages with `cow eats nuts', $364$ pages with `cow eats fish' and $29$ pages with `cat eats nuts'. This gives $P(A_1,B'_1)=0.7247$, $P(A_2,B'_2)=0.0014$, $P(A_1, B'_2)=0.2537$ and $P(A_2,B'_1)=0.0202$ and $E(A,B')=0.4523$. For the coincidence experiment $A'B'$ we find $3$ pages that contain `horse eats fish', $423$ pages with `squirrel eats nuts', $9$ pages with `horse eats nuts' and $2$ pages with `squirrel eats fish'. This gives $P(A'_1,B'_1)=0.0069$, $P(A'_2,B'_2)=0.9680$, $P(A'_1, B'_2)=0.0206$ and $P(A'_2,B'_1)=0.0046$ and $E(A',B')=0.9497$. For the expression appearing in the Clauser-Horne-Shimony-Holt variant of Bell's inequalities we get
\begin{equation}
E(A'B')+E(A'B)+E(AB')-E(AB)=2.8614
\end{equation}
which is manifestly greater than 2, and hence constitutes a strong violation of Bell's inequalities. Since googled data slightly vary with time because of the continuous incorporation of new webpages into the Google database, we should say that the data we used were collected on May 24, 2009.

Before we present our analysis, we consider two other situations involving Bell's inequalities. For the first one, instead of a sentence part such as `cat eats grass', we just consider the pair of concepts `cat' and `grass', and google the number of pages containing this pair of concepts. This gives for the pair `cat' and `grass' the number 752,000, for the pair `cow' and `meat' we find 1,240,000, for the pair `cat' and `meat' this gives 13,400,000 and for the pair `cow' and `grass' we get 7,580,000. Hence this time on a totality of 752,000+1,240,000+13,400,000+7,580,000=22,972,000 we get the fractions 752,000, 1,240,000, 13,400,000 and 7,580,000 for the various pairs considered. Again we suppose that each page is elected with equal probability, which allows us to calculate the probability for one pair to be elected. This gives $P(A_1,B_1)=752,000/22,972,000=0.0326$ for the pair `cat' and `grass', $P(A_2,B_2)=1,290,000/22,972,000=0.0540$ for the pair `cow' and `meat', $P(A_1,B_2)=13,400,000/22,972,000=0.5834$ for the pair `cat' and `meat' and $P(A_2,B_1)=7,580,000/22,972,000=0.3300$ for the pair `cow' and `grass'. In this situation we should in fact take into account that there are pages containing different of these pairs. But a more complicated calculation of the probabilities does not affect the outcome, i.e. the violation of Bell's inequalities. Knowing these probabilities, we can calculate the expectation value for this coincidence experiment by means of the equation $E(A,B)=P(A_1,B_1)+P(A_2,B_2)-P(A_2,B_1)-P(A_1,B_2)=-0.8269$. We calculate the expectation values $E(A',B)$, $E(A,B')$ and $E(A',B')$ in an analogous way making use of the Google results. For the coincidence experiment $A'B$ we find $12,500,000$ pages that contain `horse' and `grass', $1,370,000$ pages with `squirrel' and `meat', $2,270,000$ pages with `horse' and `meat' and $2,970,000$ pages with `squirrel' and `grass'. This gives $P(A'_1,B_1)=0.6541$, $P(A'_2,B_2)=0.0717$, $P(A'_1, B_2)=0.1188$ and $P(A'_2,B_1)=0.1554$ and $E(A',B)=0.4516$. For the coincidence experiment $AB'$ we find $25,100,000$ pages that contain `cat' and `fish', $3,370,000$ pages with `cow' and `nuts', $2,180,000$ pages with `cow' and `fish' and $7,070,000$ pages with `cat' and `nuts'. This gives $P(A_1,B'_1)=0.6654$, $P(A_2,B'_2)=0.0893$, $P(A_1, B'_2)=0.0578$ and $P(A_2,B'_1)=0.1874$ and $E(A,B')=0.5095$. For the coincidence experiment $A'B'$ we find $12,500,000$ pages that contain `horse' and `fish', $611,000$ pages with `squirrel'  and `nuts', $5,680,000$ pages with `horse' and `nuts' and $1,690,000$ pages with `squirrel' and 'fish'. This gives $P(A'_1,B'_1)=0.6103$, $P(A'_2,B'_2)=0.0298$, $P(A'_1, B'_2)=0.2773$ and $P(A'_2,B'_1)=0.0825$ and $E(A',B')=0.2803$. For the expression appearing in the Clauser-Horne-Shimony-Holt variant of Bell's inequalities we get
\begin{equation}
E(A'B')+E(A'B)+E(AB')-E(AB)=2.0680
\end{equation}
which is slightly bigger than 2. This means that again we have a violation here, although it is not near as strong as in the case of the conceptual combination {\it The Animal eats the Food}.

The second alternative situation we consider is the following. We suppose that there are two separated sources of knowledge. Since we do not have two separated World Wide Webs available, we use it for both sources. As we will see, this is no problem for what we want to show. Consider the experiment $A$, for {\it Cat} and {\it Grass}. We now choose one page from one of the sources of knowledge that contains `cat', and in parallel choose one page from the second source of knowledge that contains `grass', with the combination of these two pages being considered as the page that contains the couple `cat' and `grass'. Again we can calculate the probabilities and expectation values. However, this time we have to proceed as follows. We search in Google and find 98,000,000 pages containing `cat' and 68,200,000 pages containing `cow'. This means that the probability for a page from the first source of knowledge to contain `cat' is given by $P(A_1)=98,000,000/(98,000,000+68,200,000)=0.5897$, and the probability for such a page to contain `cow' is given by $P(A_2)=68,200,000/(98,000,000+68,200,000)=0.4103$. Analogously, the probability for a page from the second source of knowledge to contain `grass' is $P(B_1)=90,900,000/(90,900,000+116,000,000)=0.4393$, since 90,900,000 is the number of pages found in Google that contain `grass', and the probability for a page from the second source of knowledge to contain `meat' is $P(B_2)=116,000,000/(90,900,000+116,000,000)=0.5607$,  since 116,000,000 is the number of pages that contains `meat'. Since a page contains the pair `cat' and `grass' if it is the combined page of one page containing `cat' from the first source of knowledge and a second page containing `grass' from the second source of knowledge, it follows that the probability for this to take place is $P(A_1,B_1)=P(A_1)P(B_1)=0.2591$. Analogously, we find $P(A_2,B_2)=P(A_2)P(B_2)=0.2301$, $P(A_1,B_2)=P(A_1)P(B_2)=0.3306$ and $P(A_2,B_1)=P(A_2)P(B_1)=0.1803$. This gives $E(A,B)=-0.0218$. We calculate $E(A',B)$, $E(A,B')$ and $E(A',B')$ in an analogous way. The number of pages containing `horse' is 227,000,000, the number of pages containing `squirrel' is 28,200,000, the number of pages containing `fish' is 291,000,000 and the number of pages containing `nuts' is 60,500,000. This gives $P(A'_1)=0.8895$, $P(A'_2)=0.1105$, $P(B'_1)=0.8279$ and $P(B'_2)=0.1721$. From this it follows that $P(A'_1,B_1)=P(A'_1)P(B_1)=0.3908$, $P(A'_2,B_2)=P(A'_2)P(B_2)=0.0620$, $P(A'_1,B_2)=P(A'_1)P(B_2)=0.4987$ and $P(A'_2,B_1)=P(A'_2)P(B_1)=0.0485$, and as a consequence we have $E(A'B)=-0.0945$. Hence, in an analogous way, we get $E(A,B')=0.1176$ and $E(A',B')=0.5108$. For the expression appearing in the Clauser-Horne-Shimony-Holt variant of Bell's inequalities, this gives
\begin{equation}
E(A'B')+E(A'B)+E(AB')-E(AB)=0.5557
\end{equation}
which is very different from both previous results and does not violate Bell's inequalities. 

The reason that Bell's inequalities are not violated in this case is structural and not coincidental. Let us show this making use of the following lemma.

\bigskip
\noindent
{\bf Lemma}:{
If $x$, $x'$, $y$ and $y'$ are real numbers such that $-1\le x, x', y , y\le +1$ and $S=xy + xy' + x'y - x'y'$ then $-2\le S \le +2$.}

\bigskip
\noindent
Proof: Since $S$ is linear in all four variables $x$, $x'$, $y$, $y'$, it must take on its maximum and minimum values at the corners of the domain of this quadruple of variables, that is, where each of $x$, $x'$, $y$, $y'$ is +1 or -1. Hence at these corners $S$ can only be an integer between -4 and +4. But $S$ can be rewritten as $(x + x')(y + y') - 2x'y'$, and the two quantities in parentheses can only be 0, 2, or -2, while the last term can only be -2 or +2, so that S cannot equal -3, +3, -4, or +4 at the corners. 

\bigskip
\noindent
Since in the situation considered we have $P(A_i,B_j)=P(A_i)P(B_j)$, $P(A'_i,B_j)=P(A'_i)P(B_j)$, $P(A_i,B'_j)=P(A_i)P(B'_j)$ and $P(A'_i,B'_j)=P(A'_i)P(B'_j)$, we have $E(A,B)=E(A)E(B)$, $E(A',B)=E(A')E(B)$, $E(A,B')=E(A)E(B')$ and $E(A',B')=E(A')E(B')$, and hence from the lemma it follows that
\begin{eqnarray}
&&-2\le E(A'B')+E(A'B)+E(AB')-E(AB) \le +2 
\end{eqnarray}
which proves the Clauser-Horne-Shimony-Holt variant of Bell's inequalities to be valid.

The foregoing examples and analysis show that the non-product nature of probabilities $P(A_i,B_j)$, $P(A'_i,B_j)$, $P(A_i,B'_j)$ and $P(A'_i,B'_j)$ is key to the violation of Bell's inequalities, for example $P(A_i,B_j)\not=P(A_i)P(B_j)$. If we understand why these coincidence probabilities are not of the product nature we can also understand why Bell's inequalities are violated in the situations of consideration. The answer is simple, in fact, and already implied in the above analysis, but let us make it explicit. Consider for example $P(A_2,B_2)$ and let us analyze why it is different from $P(A_2)P(B_2)$. We have that $P(A_2,B_2)$ is the probability for a random page from the World Wide Web to contain the sentence part `cow eats meat' in our first example, and then we find $P(A_2,B_2)=0.0717$, or to contain the pair of concepts `cow' and `meat' in our second example, and then we find $P(A_2,B_2)=0.0540$. While $P(A_2)P(B_2)$ is the probability that, for two pages chosen at random, one contains `cow' and the other contains `meat', and then we find $P(A_2)P(B_2)=0.2301$. This value is very different from the other two, and we can easily see why this is the case. The probability of finding the sentence part `cow eats meat' is low, because of its very meaning, cows not being in the habit of eating meat. Therefore, the number of webpages containing both `cow' and `meat' is very low, so that the probability for any such page to be chosen is very low too. If however two `separated' or `independent' pages are chosen at random, the probability for the one to contain `cow' and the other to contain `meat' is substantial and not small. The fundamental reason for this difference is the fact that in the latter case the pages are separated or independent, not connected in meaning. Indeed, it is because a webpage naturally contains concepts that are all interrelated in meaning that the co-occurrence of `cow' and `meat' in one webpage is so small. In the next section we will analyze this situation in further detail.

\subsection{Meaning and Coherence}
Before that, however, we want to show that the violation of Bell's inequalities -- and hence the presence of entanglement -- is based on exactly the same mathematical structure for quantum entities and concept combinations. For quantum entities, entanglement is mathematically structured because the wave function $\psi(x,y)$ of a joint quantum entity of two sub-entities is not necessarily a product $\psi_1(x)\psi_2(y)$ of the wave function $\psi_1(x)$ of one of the sub-entities and the wave function $\psi_2(y)$ of the other sub-entity. Let us show that this is what also happens for concepts when they are combined. As said, human or artificial memory structures relate to concepts like ordinary matter -- hence also measuring apparatuses -- relates to quantum particles. Three-dimensional space is considered to be the theatre of macroscopic material entities, i.e. the collection of `locations' where such macroscopic material entities can be situated, and also the medium through which quantum particles communicate with macroscopic material entities, treated as measurement apparatuses in the formalism of quantum mechanics.

The items {\it Cat}, {\it Cow}, {\it Horse} and {\it Squirrel} are exemplars of the concept {\it Animal} and the items {\it Grass}, {\it Meat}, {\it Fish} and {\it Nuts} are exemplars of the concept {\it Food}. There are many more exemplars of both concepts. Let us consider some, for instance {\it Bear}, {\it Dog}, {\it Fish}, {\it Bird}, etc\ldots and some exemplars of {\it Food}, such as {\it Fruits}, {\it Milk}, {\it Vegetables}, {\it Potatos}, etc \ldots. For the specific type of measurement that we have considered with respect to Bell's inequalities -- responding to the question `which is a good example of' -- we can consider the different exemplars of both concepts as points in the memory structure of a human mind, and denote them $x_1, x_2, \ldots, x_n$ for {\it Animal} and $y_1, y_2, \ldots, y_m$ for {\it Food}. {\it Animal} can then be written as a wave function $\psi_1(x)$, where $x$ can take the values $x_1, x_2, \ldots, x_n$, and $|\psi_1(x_i)|^2$ is the weight of the exemplar $x_i$ for the concept {\it Animal} with respect to the measurement `is a good example of {\it Animal}'. Likewise, the concept {\it Food} can be represented by the wave function $\psi_2(y)$, where $y$ takes the values $y_1, y_2, \ldots, y_m$ and $|\psi(y_j)|^2$ is the weight of exemplar $y_j$ for the concept {\it Food} with respect to the measurement `is a good example of {\it Food}'. If the concepts {\it Animal} and {\it Food} form a combination of concepts --{\it The Animal eats the Food}, for example -- we can describe this by means of a wave function $\psi(x,y)$, where $|\psi(x_i,y_j)|^2$ is the weight of the couple of exemplars $x_i$ and $y_j$ for the considered combination of the concepts {\it Animal} and {\it Food} with respect to the measurement `is a good example of {\it The Animal eats the Food}'. The foregoing analysis shows exactly that $\psi(x,y)$ is {\it not} a product. Indeed, if it was, Bell's inequalities would not be violated. The reason why it is not a product is clearly illustrated by the above combination of {\it Animal} and {\it Food} into {\it The Animal eats the Food}. This combination introduces a wave function that attributes weights to couples of exemplars $(x_i, y_j)$ in a new way, i.e. different from how weights are attributed by component wave functions describing the measurements related to the questions `which is a good example of {\it Animal}' and `which is a good example of {\it Food}' separately, and a product of such component wave functions. This is because {\it The Animal eats the Food} is not only a combination of concepts, but a new concept on its own account.  It is this new concept that determines the values attributed to weights of couples of exemplars, which will therefore be different from the values attributed if we consider only the products of weights determined by the constituent concepts. Likewise, it is clear that `all functions of two variables', i.e. all functions $\psi(x,y)$, will be possible expressions of states of this new concept. Indeed, all types of combinations, e.g. {\it The Animal dislikes the Food}, or {\it Look how this Animal tries to eat this Food}, or {\it I do not think that this Animal will eat this Food}, etc\ldots, will introduce different states $\psi(x,y)$ which are wave functions in the product space $\{x_1, \ldots, x_n\}\times\{y_1,\ldots,y_m\}$. This shows that combining concepts in a natural and understandable way gives rise to entanglement, and it does so structurally in a completely analogous way as entanglement appears in quantum mechanics, namely by allowing all functions of joint variables of two entities to play a role as wave functions describing states of the joint entity consisting of these two entities.

Let us reflect more carefully on why this is the way concepts combine. Consider the conceptual combination {\it The Animal eats the Food}. When we search the World Wide Web for phrases that are exemplars of this conceptual combination, calculating their probability of appearance, we find that Bell's inequalities are violated. Why? Because the meaning of the conceptual combination {\it The Animal eats the Food} is reflected in the phrases that are exemplars of this conceptual combination. And this holds for all texts on the World Wide Web, which tend to be meaningful. We find an abundance of exemplars of `cow eats grass' and very little exemplars of `squirrel eats grass' on the World Wide Web, because the concepts {\it Animal} and {\it Food} combine with each other to form {\it The Animal eats the Food} throughout the meaning that these concepts carry. There is a property that is often referred to in discussions about quantum entities  \cite{albrecht1994} and that plays a very similar role to that played by meaning for concepts and their combinations. This property is called {\it coherence}. It is exactly when coherence governs for a quantum entity or different quantum entities that Bell's inequalities can be violated. Coherence, like meaning for concepts, is a property that exists between quantum entities, a priori to localization of such quantum entities. Coherent quantum entities give rise to non-locality because the correlations produced by coherence are a priori to the experimental detection of the consequences of these correlations in localized states of such quantum entities. The same applies to concepts. The correlations carried by meaning are a priori to their detection in exemplars of the concepts, in line with what we found on the World Wide Web for the combination of concepts {\it The Animal eats the Food}.

To illustrate this in more concrete terms, let us consider a dictionary containing all exemplars of {\it Animal} and all exemplars of {\it Food} ordered alphabetically. Let us consider again the example of section \ref{violationbellsinequalities}. A choice is made between two exemplars of {\it Animal}, namely {\it Cat} and {\it Cow}, and let us suppose this is done by a person who uses a pencil in his or her left hand to do so. A choice is also made between two exemplars of {\it Food}, namely {\it Grass} and {\it Meat}, and this is done by the same person, who this time uses a pencil in his or her right hand. They are asked to make their choice by answering the question of `which is a good example of' for the combination of concepts {\it The Animal eats the Food}. Exactly like we analyzed in section \ref{violationbellsinequalities}, there will be correlations between the chosen concepts that violate Bell's inequalities. If the left hand indicates {\it Cow}, it is substantially more probable for the right hand to indicate {\it Grass} than {\it Meat}, and similarly, if the pencil in the left hand points to {\it Cat}, it is more probable for the right hand pencil to point to {\it Meat} instead of to {\it Grass}. Using the same setup for the other coincidence experiments considered in section \ref{violationbellsinequalities}, i.e. including {\it Horse} and {\it Squirrel} and {\it Fish} and {\it Nuts}, and carrying out experiments on the necessary combination, we find that Bell's inequalities are violated based on a similar collection of data to that considered  in section \ref{violationbellsinequalities}. This indicates the presence of non-locality, which becomes more apparent if we make the following change to the experimental situation. We now consider two dictionaries, placed on either side of a table. The test person uses his or her left hand to choose between {\it Cat} and {\it Cow} in the left dictionary, and his or her right hand to choose between {\it Grass} and {\it Meat} in the right dictionary. Two other persons present only watch how the words are chosen from each of the dictionaries. They will have a subjective experience of the non-locality when they compare the outcomes of {\it Cat} in one dictionary being correlated with {\it Meat} in the other dictionary and {\it Cow} in one dictionary being correlated with {\it Grass} in the other. If the two onlookers are familiar with the procedure of the experiment, i.e. that the choosing is done by somebody faced with two dictionaries on either side of a table, and, more importantly, that the choosing is governed by his or her efforts to find meaningful illustrations of the combination of concepts {\it The Animal eats the Food}, corresponding to exemplars of this combination that yield `good examples', they would not classify the non-locality as mysterious. Indeed, the non-local correlations originate from the fact that the test person does the choosing based on the meaning of the sentence {\it The Animal eats the Food}. Hence there is no message going from one dictionary to the other producing the correlations. 

We can see that this is exactly how Bell's inequalities are violated by entangled quantum entities too. Here the violation is not originated by `connection through meaning' but by `connection through coherence' -- which is just another way of saying `entangled'. To be more specific about this analogy, we will give the example of an archetypical type of pair of entangled quantum entities, namely two quantum particles with entangled spins 1/2. The spin state of such two quantum particles is described by the singlet state
\begin{equation}
\psi_S={1 \over \sqrt{2}}(|up\rangle\otimes|down\rangle-|down\rangle\otimes|up\rangle)
\end{equation}
Our interpretation of this state is that it is not a state of spin zero, but a state with no spin. None of the two particles flying apart has already a spin. There is just the quantum conceptual content carried by the quantum coherence that `spin needs to be zero whenever it is forced to appear'. In exact analogy with the conceptual content of the sentence {\it The Animal eats the Food}, if {\it Animal} equals {\it Cow}, then {\it Food} needs to be {\it Grass}, while if {\it Animal} equals {\it Cat}, then {\it Food} must be {\it Meat}. Even though the correlation as experimented in section \ref{violationbellsinequalities} will be less strict than in the case of the spins, it is still sufficiently significant to lead to a violation of Bell's inequalities. In the context of the new interpretation that we propose, this gives us reason to assume that when one of the two spins is forced to be up, the other one will simultaneously be forced to be down, and vice versa. This produces correlations that violate Bell's inequalities, {\it because} these correlations arise during a measurement that is taking place from a state that carries a quantum conceptual content that does not refer to specific values of the spin, but only says that `if spin is created then its total value needs to be zero' on the conceptual level. The suggestion that this is what takes place in quantum mechanics is even indicated by the form of the mathematical expression of the singlet state. Indeed, we know that the singlet state is the anti-symmetric sum for couples of spin, whatever their direction. In fact, it shows us mathematically that it is a state where the directions of spins have not yet been created. This must be so, because otherwise it could not be the anti-symmetric superposition for any spins chosen in the component product states.

The following is intended to forestall comments that there is a big difference between, on the one hand, dictionaries connected by means of a sentient human being and their knowledge of the meaning of a sentence causing the correlations to violate Bell's inequalities, and on the other, entangled spins connecting measuring apparatus and their quantum coherence -- or entanglement -- doing the same. As we demonstrated many years ago, Bell's inequalities can be violated in a very similar way by ordinary macroscopic material entities, i.e. without the presence of a sentient being connecting the measuring apparatuses. The first example we worked out for this purpose was that of Bell's inequalities being violated by two vessels of water interconnected by a tube \cite{aerts1982,aerts1985a,aerts1985b}. Later, we elaborated versions of the same idea leading to violations of Bell's inequalities with equal numerical values as those predicted by quantum mechanics \cite{aerts1991,aertsaertsbroekaertgabora2000}. In more recent years, we studied the macroscopic model in detail to work out a `geometrical representation of entanglement as internal constraint' \cite{aertsdhondtdhooghe2005}. We will briefly present the original `vessels of water model', which should suffice to make our point for the present article.

We consider two vessels $V_A$ and $V_B$ interconnected by a tube $T$, each of them containing 10 liters of transparent water. Coincidence experiments $A$ and $A'$ consist in siphons $S_A$ and $S_B$ pouring out water from vessels $V_A$ and $V_B$, respectively, and collecting the water in reference vessels $R_A$ and $R_B$, where the volume of collected water is measured, as shown in Figure 1. If more than 10 liters is collected for experiments $A$ or $B$ we put $E(A)=+1$ or $E(B)=+1$, respectively, and if less than 10 liters is collected for experiments $A$ or $B$, we put $E(A)=-1$ or $E(B)=-1$, respectively. We define experiments $A'$ and $B'$, which consist in taking a small spoonful of water out of the left vessel and the right vessel, respectively, and verifying whether the water is transparent. We have $E(A')=+1$ or $E(A')=-1$, depending on whether the water in the left vessel turns out to be transparent or not, and $E(B')=+1$ or $E(B')=-1$ depending on whether the water in the right vessel turns out to be transparent or not. We define $E(AB)=+1$ if $E(A)=+1$ and $E(B)=+1$ or $E(A)=-1$ and $E(B)=-1$, and $E(A,B)=-1$ if $E(A)=+1$ and $E(B)=-1$ or $E(A)=-1$ and $E(B)=+1$, if the coincidence experiment $AB$ is performed. Note that we follow the traditional way in which the expectation value of the coincidence experiments $AB$ is defined. In a similar way, we define $E(A'B)$, $E(AB')$ and $E(A'B')$, the expectation value corresponding to the coincidence experiments $A'B$, $AB'$ and $A'B'$, respectively. 
\begin{figure}[H]
\centerline {\includegraphics[width=11cm]{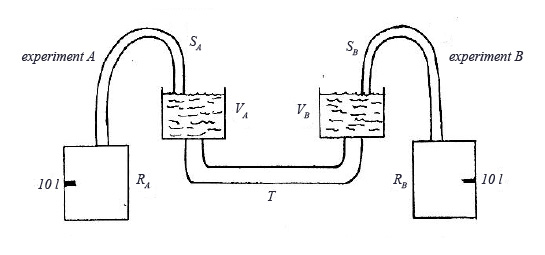}}
\caption{Two vessels $V_A$ and $V_B$ containing 10 liters of transparent water each are interconnected by a tube $T$. Coincidence experiments $A$ and $B$ consist in siphons $S_A$ and $S_B$ pouring out water from vessels $V_A$ and $V_B$, respectively, and collecting the water in reference vessels $R_A$ and $R_B$, where the volume of collected water is measured.}
\end{figure}
\noindent
Hence, concretely $E(A'B)=+1$ if $E(A')=+1$ and $E(B)=+1$ or $E(A')=-1$ and $E(B)=-1$ and the coincidence experiment $A'B$ is performed. And we have $E(AB')=+1$ if $E(A)=+1$ and $E(B')=+1$ or $E(A)=-1$ and $E(B')=-1$ and the coincidence experiment $AB'$ is performed, and further $E(A'B')=+1$ if $E(A')=+1$ and $E(B')=+1$ or $E(A')=-1$ and $E(B')=-1$ and the coincidence experiment $A'B'$ is performed. Since each vessel contains 10 liters of transparent water, we find experimentally that $E(AB)=-1$, $E(A'B)=+1$, $E(AB')=+1$ and $E(A'B')=+1$, which gives
\begin{equation}
E(A',B')+E(A'B)+E(AB')-E(AB)=+4
\end{equation}
This is the maximum possible violation of Bell's inequalities.

Let us analyze this example in the light of the foregoing sections of this article. The main reason why these interconnected water vessels can violate Bell's inequalities is because the water in the vessels has not yet been subdivided into two volumes before the measurement starts. The water in the vessels is only `potentially' subdivided into volumes of which the sum is 20 liters. It is not until the measurement is actually carried out that one of these potential subdivisions actualizes, i.e. one part of the 20 liters is collected in reference vessel $R_A$ and the other part is collected in reference vessel $R_B$. This is very similar to the combinations of concepts {\it The Animal eats the Food} not being collapsed into one of the four possibilities {\it The Cat eats the Grass}, {\it The Cat eats the Meat}, {\it The Cow eats the Grass} or {\it The Cow eats the Meat} before the coincidence measurement $AB$ starts. It is the coincidence measurement itself which makes the combination of concepts {\it The Animal eats the Food} collapse into one of the four possibilities. The same holds for the interconnected water vessels. The coincidence measurement $AB$ with the siphons is what causes the total volume of 20 liters of water to be split into two volumes, and it is this which creates the correlation for $AB$ giving rise to $E(AB)=-1$. We believe that a similar process takes place for the typical quantum mechanical experiments violating Bell's inequalities. For example, for the two spin 1/2 quantum particles in the singlet spin state, even if these particles fly apart, they are not yet two particles with opposite spins flying apart. There are two particles flying apart, and their spins are being created during the correlation experiments, which then leads to a violation of Bell's inequalities. It is the presence of `coherence' within the entangled quantum particles between the measurement apparatuses that is necessary to be able to provoke a violation of Bell's inequalities. Likewise, it is the presence of `meaning' between the two dictionaries, in the mind of the person choosing the words with his or her right hand and left hand from each of the dictionaries, that is necessary to provoke a violation of Bell's inequalities. And again, it is the presence of `water' in the two connected vessels which is necessary to provoke a violation of Bell's inequalities.

What is so special about concepts that we decided to propose this new quantum interpretation based on them? Water, as well as other macroscopic mechanisms that we have investigated with the aim of understanding more of the nature of quantum coherence \cite{aerts1991,aertsaertsbroekaertgabora2000,aertsdhondtdhooghe2005}, has not allowed us to build models for more than two entangled entities. Concepts entangle in exactly the same way as quantum systems do, also for more than two entities. Partly this is because concepts and their combinations do not need to be `inside space', unlike water and the mechanical systems built in \cite{aerts1991,aertsaertsbroekaertgabora2000,aertsdhondtdhooghe2005}, and also waves. The hypothesis put forward in this article and in \cite{aerts2009a}, namely the idea that quantum entities behave like conceptual entities, is therefore compatible with a view that we had already put forward in earlier work, viz. that non-locality means non-spatiality. Space, which could be three dimensional and Euclidean or four dimensional and Minkowskian or curved by gravity -- in which case it is often called space-time -- should not be considered as the overall theatre of reality, but rather as `a space' that has grown together with how macroscopic material entities have grown out of the micro-world. Hence, space is `the space of macroscopic material entities', while quantum entities are most of the time not inside `this' space \cite{aerts1999,aerts1998a,aerts1998b,aerts1995,aerts1994,aerts1992,aerts1986}.

If we put forward the hypothesis that `quantum entities are the conceptual entities exchanging (quantum) meaning -- identified as quantum coherence -- between measuring apparatuses, and more generally between entities made of ordinary matter', it might seem as if we want to develop a drastic antropomorphic view about what goes on in the micro-world. It could give the impression that in the view we develop `what happens in our macro-world, namely people using concepts and their combinations to communicate' already took place in the micro-realm too, namely `measuring apparatuses, and more generally entities made of ordinary matter, communicate with each other and the words and sentences of their language of communication are the quantum entities and their combinations'. This is certainly a fascinating and eventually also possible way to develop a metaphysics compatible with the explanatory framework that we put forward. However, such a metaphysics it is not a necessary consequence of our basic hypothesis, and only further detailed research can start to see which aspects of such a drastic metaphysical view formulated above are eventually true and which are not at all. We also do not have to exclude eventual fascinating metaphysical speculations related to this new interpretation and explanatory framework from the start. An open, but critical and scientific attitude is what is most at place with respect to this aspect of our approach, and this is what we will attempt in the future. In this sense we want to mention a possibility showing that our basic hypothesis can also lead to a much less antropomorphic view.

If instead of `conceptual entity' we use the notion of `sign', we can formulate our basic hypothesis in much less antropomorphic terms as follows: `Quantum entities are signs exchanged between measuring apparatuses and more generally between entities made of ordinary matter'. We use the notion of `sign' here as it was generally introduced into semiotics \cite{chandler2002}. Semiotics is the study of the exchange of signs of any type, which means that it covers animal communication, but also the exchange of signs, including icons, between computer interfaces. This much less antropomorphic view of quantum entities as signs instead of cognitive entities allows to interpret measurement apparatuses and more general entities made of ordinary matter as interfaces for these signs instead of memories for conceptual entities. However, what is a fundamental consequence of our basic hypothesis, whether we consider its `cognitive version' or its  `semiotic version', is that communication of some type takes place, and, more specifically, that the language or system of signs used in this communication evolved symbiotically with the memories for this language or the interfaces for these signs. This introduces in any case a radically new way to look upon the evolution of the part of the universe we live in, namely the part of the universe consisting of entities of ordinary matter and quantum fields. Any mechanistic view, whether the mechanistic entities are conceived of as particles, as waves or as both, cannot work out well if the reality is one of co-evolving concepts and memories or signs and interfaces.

\section{Interference and Superposition} \label{interferencesuperposition}

Next to entanglement there is another fundamental aspect of quantum entities and how they interact, namely interference. In this section we analyze the role played by the phenomenon of interference in our new interpretation and how our explanatory framework provides a simple explanation for its appearance. To this end, we will work out in detail a specific example of how concepts interfere.

\subsection{Fruits interfering with Vegetables}
We consider the concepts {\it Fruits} and {\it Vegetables}, two exemplars of the concept {\it Food}. And we consider a collection of exemplars of {\it Food}, more specifically those listed in Table 1. Then we consider the following experimental situation: Human beings -- `subjects', in the terminology of psychology -- are asked to respond to the following three elements: {\it Question $A$}: `Choose one of the exemplars from the list of Table 1 that you find a good example of {\it Fruits}'. {\it Question $B$}: `Choose one of the exemplars from the list of Table 1 that you find a good example of {\it Vegetables}'. {\it Question $A$ or $B$}: `Choose one of the exemplars from the list of Table 1 that you find a good example of {\it Fruits or Vegetables}'. Then we calculate the relative frequency $\mu(A)_k$, $\mu(B)_k$ and $\mu(A\ {\rm or}\ B)_k$, i.e the number of times that exemplar $k$ is chosen divided by the total number of choices made in response to the three questions $A$, $B$ and $A\ {\rm or}\ B$, respectively, and interpret this as an estimate for the probabilities that exemplar $k$ is chosen for questions $A$, $B$ and $A\ {\rm or}\ B$, respectively.
These relative frequencies are given in Table 1. For example, for {\it Question $A$}, from 10,000 subjects, 359 chose {\it Almond}, hence $\mu(A)_1=0.0359$, 425 chose {\it Acorn}, hence $\mu(A)_2=0.0425$, 372 chose {\it Peanut}, hence $\mu(A)_3=0.0372$, $\ldots$, and 127 chose {\it Black Pepper}, hence $\mu(A)_{24}=0.0127$. Analogously for {\it Question $B$}, from 10,000 subjects, 133 chose {\it Almond}, hence $\mu(B)_1=0.0133$, 108 chose {\it Acorn}, hence $\mu(B)_2=0.0108$, 220 chose {\it Peanut}, hence $\mu(B)_3=0.0220$, $\ldots$, and 294 chose {\it Black Pepper}, hence $\mu(B)_{24}=0.0294$, and for {\it Question $A\ {\rm or}\ B$}, 269 chose {\it Almond}, hence $\mu(A\ {\rm or}\ B)_1=0.0269$, 249 chose {\it Acorn}, hence $\mu(A\ {\rm or}\ B)_2=0.249$, 269 chose {\it Peanut}, hence $\mu(A\ {\rm or}\ B)_3=0.269$, $\ldots$, and 222 chose {\it Black Pepper}, hence $\mu(A\ {\rm or}\ B)_{24}=0.222$.
\begin{table}[htb]
\footnotesize
\begin{center}
\begin{tabular}{|llllllll|}
\hline 
\multicolumn{2}{|l}{} & \multicolumn{1}{l}{$\mu(A)_k$} & \multicolumn{1}{l}{$\mu(B)_k$} & \multicolumn{1}{l}{$\mu(A\ {\rm or}\ B)_k$} & \multicolumn{1}{l}{${\mu(A)_k+\mu(B)_k \over 2}$} & \multicolumn{1}{l}{$\lambda_k$} & \multicolumn{1}{l|}{$\theta$} \\
\hline
\multicolumn{8}{|l|}{\it $A$=Fruits, $B$=Vegetables} \\
\hline
1 & {\it Almond} & 0.0359 & 0.0133 & 0.0269 & 0.0246 & 0.0218 & 83.8854$^\circ$ \\
2 & {\it Acorn} & 0.0425 & 0.0108 & 0.0249 & 0.0266 & -0.0214 & -94.5520$^\circ$ \\
3 & {\it Peanut} & 0.0372 & 0.0220 & 0.0269 & 0.0296 & -0.0285 & -95.3620$^\circ$ \\
4 & {\it Olive} & 0.0586 & 0.0269 & 0.0415 & 0.0428 & 0.0397 & 91.8715$^\circ$ \\
5 & {\it Coconut} & 0.0755 & 0.0125 & 0.0604 & 0.0440 & 0.0261 & 57.9533$^\circ$ \\
6 & {\it Raisin} & 0.1026 & 0.0170 & 0.0555 & 0.0598 & 0.0415 & 95.8648$^\circ$ \\
7 & {\it Elderberry} & 0.1138 & 0.0170 & 0.0480 & 0.0654 & -0.0404 & -113.2431$^\circ$ \\ 
8 & {\it Apple} & 0.1184 & 0.0155 & 0.0688 & 0.0670 & 0.0428 & 87.6039$^\circ$ \\ 
9 & {\it Mustard} & 0.0149 & 0.0250 & 0.0146 & 0.0199 & -0.0186 & -105.9806$^\circ$ \\
10 & {\it Wheat} & 0.0136 & 0.0255 & 0.0165 & 0.0195 & 0.0183 & 99.3810$^\circ$ \\ 
11 & {\it Root Ginger} & 0.0157 & 0.0323 & 0.0385 & 0.0240 & 0.0173 & 50.0889$^\circ$ \\
12 & {\it Chili Pepper} & 0.0167 & 0.0446 & 0.0323 & 0.0306 & -0.0272 &  -86.4374$^\circ$ \\ 
13 & {\it Garlic} & 0.0100 & 0.0301 & 0.0293 & 0.0200 & -0.0147 & -57.6399$^\circ$ \\
14 & {\it Mushroom} & 0.0140 & 0.0545 & 0.0604 & 0.0342 & 0.0088 & 18.6744$^\circ$ \\
15 & {\it Watercress} & 0.0112 & 0.0658 & 0.0482 & 0.0385 & -0.0254 &  -69.0705$^\circ$ \\
16 & {\it Lentils} & 0.0095 & 0.0713 & 0.0338 & 0.0404 & 0.0252 & 104.7126$^\circ$ \\
17 & {\it Green Pepper} & 0.0324 & 0.0788 & 0.0506 & 0.0556 & -0.0503 & -95.6518$^\circ$ \\
18 & {\it Yam} & 0.0533 & 0.0724 & 0.0541 & 0.0628 & 0.0615 & 98.0833$^\circ$ \\
19 & {\it Tomato} & 0.0881 & 0.0679 & 0.0688 & 0.0780 & 0.0768 & 100.7557$^\circ$ \\
20 & {\it Pumpkin} & 0.0797 & 0.0713 & 0.0579 & 0.0755 & -0.0733 & -103.4804$^\circ$  \\
21 & {\it Broccoli} & 0.0143 & 0.1284 & 0.0642 & 0.0713 & -0.0422 & -99.6048$^\circ$ \\
22 & {\it Rice} & 0.0140 & 0.0412 & 0.0248 & 0.0276 & -0.0238 & -96.6635$^\circ$ \\ 
23 & {\it Parsley} & 0.0155 & 0.0266 & 0.0308 & 0.0210 & -0.0178 & -61.1698$^\circ$ \\
24 & {\it Black Pepper} & 0.0127 & 0.0294 & 0.0222 & 0.0211 & 0.0193 & 86.6308$^\circ$ \\
\hline
\end{tabular}
\end{center}
\caption{Interference data for concepts {\it A=Fruits} and {\it B=Vegetables}. The probability of a person choosing one of the exemplars as an example of {\it Fruits} (and as an example of {\it Vegetables}, respectively), is given by $\mu(A)$ (and $\mu(B)$, respectively) for each of the exemplars. The probability of a person choosing one of the exemplars as an example of {\it Fruits or Vegetables} is $\mu(A\ {\rm or}\ B)$ for each of the exemplars. The classical probability would be given by ${\mu(A)+\mu(B) \over 2}$, and $\theta$ is the quantum phase angle provoking the quantum interference effect.}
\end{table}

It should be noted that the data in Table 1 were not collected by actually asking the three questions to a fixed number of persons but derived from standard psychological experiments performed by James Hampton to measure the typicality of the 24 exemplars in Table 1 with respect to the concepts {\it Fruits}, {\it Vegetables} and their disjunction `{\it Fruits or Vegetables}' \cite{hampton1988}. Hampton asked the subjects to `estimate the typicality of the different exemplars with respect to the concepts {\it Fruits}, {\it Vegetables} and {\it Fruits or Vegetables}'. However, this set of data is appropriate for our experiment since the estimated typicality of an exemplar is strongly correlated with the frequency with which it is chosen as `a good example'. In the practice of experimental psychology, preference is often given to measuring typicality by asking subjects to estimate it, because this approach requires a much smaller sample of subjects to find statistically relevant results -- 40 in the case of Hampton's experiment. The reason is that each subject involved in an estimation test gives 24 answers, while in a choice test they give only one. This difference is irrelevant to our proposal for a new interpretation of quantum mechanics. For further information we refer to \cite{aerts2009b}, which contains a detailed description of the calculation of the frequency data of Table 1 from Hampton's typicality data. Remark that we could also have used Google to estimate these probabilities, as we have done in section \ref{entanglement} for calculating the expectation values to be substituted in Bell's inequalities. We would then have used the number of pages a Google search indicated for the words `fruits' and `apple' and used this number of pages divided by the total number of pages in play to calculate the probability for apple to be chosen as a `good example of fruits' amongst the 24 exemplars. We decided to use data from a psychology experiment this time for two reasons. First of all because the data were available due to Hampton's experiments \cite{hampton1988}, and secondly to contrast them with the Google data based choice that we made in our analysis of entanglement in section \ref{entanglement}. 

Our analysis here of the interference phenomenon goes further than what we did on entanglement in section \ref{entanglement}, since we explicitly construct the quantum mechanical model in Hilbert space for the pair of concepts {\it Fruit} and {\it Vegetable} and their disjunction `{\it Fruit or Vegetable}', and show that quantum interference models the experimental results well. So let us proceed now by explicitly constructing a complex Hilbert space quantum system, and see how it models the experimental data gathered in \cite{hampton1988}.

We represent the measurement of `a good example of' by means of a self-adjoint operator with spectral decomposition $\{M_k\ \vert\ k=1,\ldots,24\}$ where each $M_k$ is an orthogonal projection of the Hilbert space ${\cal H}$ corresponding to item $k$ from the list of items in Table 1. The concepts {\it Fruits}, {\it Vegetables} and `{\it Fruits or Vegetables}' are represented by unit vectors $|A\rangle$, $|B\rangle$ and ${1 \over \sqrt{2}}(|A\rangle+|B\rangle)$ of the Hilbert space ${\cal H}$. Following the standard rules of quantum mechanics the probabilities, $\mu(A)_k$, $\mu(B)_k$ and $\mu(A\ {\rm or}\ B)_k$ are given by
\begin{eqnarray}
\mu(A)_k=\langle A|M_k|A\rangle \quad \mu(B)_k=\langle B|M_k|B\rangle \quad \mu(A\ {\rm or}\ B)_k={1 \over 2}\langle A+B|M_k|A+B\rangle
\end{eqnarray}
and a straightforward calculation gives
\begin{equation} \label{muAorB}
\mu(A\ {\rm or}\ B)_k={1 \over 2}(\langle A|M_k|A\rangle+\langle B|M_k|B\rangle+\langle A|M_k|B\rangle+\langle B|M_k|A\rangle)={1 \over 2}(\mu(A)_k+\mu(B)_k)+\Re\langle A|M_k|B\rangle
\end{equation}
where $\Re\langle A|M_k|B\rangle$ is the interference term. Let us introduce $|e_k\rangle$ the unit vector on $M_k|A\rangle$ and $|f_k\rangle$ the unit vector on $M_k|B\rangle$, and put $\langle e_k|f_k\rangle=c_ke^{i\gamma_k}$. Then we have $|A\rangle=\sum_{k=1}^{24}a_ke^{i\alpha_k}|e_k\rangle$ and $|B\rangle=\sum_{k=1}^{24}b_ke^{i\beta_k}|f_k\rangle$ which gives
\begin{equation} \label{ABequation}
\langle A|B\rangle=(\sum_{k=1}^{24}a_ke^{-i\alpha_k}\langle e_k|)(\sum_{l=1}^{24}b_le^{i\beta_l}|f_l\rangle) =\sum_{k=1}^{24}a_kb_kc_ke^{i(\beta_k-\alpha_k+\gamma_k)}=\sum_{k=1}^{24}a_kb_kc_ke^{i\phi_k}
\end{equation}
where we have put $\phi_k=\beta_k-\alpha_k+\gamma_k$. Further we have
\begin{eqnarray}
\mu(A)_k&=&\langle A|M_k|A\rangle=(\sum_{l=1}^{24}a_le^{-i\alpha_l}\langle e_l|)(a_ke^{i\alpha_k}|e_k\rangle)=a_k^2 \\
\mu(B)_k&=&\langle B|M_k|B\rangle=(\sum_{l=1}^{24}b_le^{-i\beta_l}\langle f_l|)(b_ke^{i\beta_k}|f_k\rangle)=b_k^2 \\
\langle A|M_k|B\rangle&=&(\sum_{l=1}^{24}a_le^{-i\alpha_l}\langle e_l|)M_k|(\sum_{m=1}^{24}b_me^{i\beta_m}|f_m\rangle)=a_kb_ke^{i(\beta_k-\alpha_k)}\langle e_k|f_k\rangle=a_kb_kc_ke^{i\phi_k}
\end{eqnarray}
which, making use of (\ref{muAorB}), gives
\begin{equation} \label{muAorBequation}
\mu(A\ {\rm or}\ B)_k={1 \over 2}(\mu(A)_k+\mu(B)_k)+c_k\sqrt{\mu(A)_k\mu(B)_k}\cos\phi_k
\end{equation}
We choose $\phi_k$ such that
\begin{equation} \label{cosequation}
\cos\phi_k={2\mu(A\ {\rm or}\ B)_k-\mu(A)_k-\mu(B)_k \over 2c_k\sqrt{\mu(A)_k\mu(B)_k}}
\end{equation}
and hence (\ref{muAorBequation}) is satisfied. We now have to determine $c_k$ in such a way that $\langle A|B\rangle=0$. Remark that from $\sum_{k=1}^{24}\mu(A\ {\rm or}\ B)_k=1$ and (\ref{muAorBequation}), and with the choice of $\cos\phi_k$ that we made in (\ref{cosequation}), it follows that $\sum_{k=1}^{24}c_k\sqrt{\mu(A)_k\mu(B)_k}\cos\phi_k=0$. Taking into account (\ref{ABequation}), which gives $\langle A|B\rangle=\sum_{k=1}^{24}a_kb_kc_k(\cos\phi_k+i\sin\phi_k)$, and making use of $\sin\phi_k=\pm\sqrt{1-\cos^2\phi_k}$, we have
\begin{eqnarray}
\langle A|B\rangle=0
&\Leftrightarrow&
\sum_{k=1}^{24}c_k\sqrt{\mu(A)_k\mu(B)_k}(\cos\phi_k+i\sin\phi_k)=0 \\
&\Leftrightarrow&\sum_{k=1}^{24}c_k\sqrt{\mu(A)_k\mu(B)_k}\sin\phi_k=0 \\ \label{conditionequation}
&\Leftrightarrow&\sum_{k=1}^{24}\pm\sqrt{c_k^2\mu(A)_k\mu(B)_k-(\mu(A\ {\rm or}\ B)_k-{\mu(A)_k+\mu(B)_k \over 2})^2}=0
\end{eqnarray}
We introduce the following quantities
\begin{equation} \label{lambdak}
\lambda_k=\pm\sqrt{\mu(A)_k\mu(B)_k-(\mu(A\ {\rm or}\ B)_k-{\mu(A)_k+\mu(B)_k \over 2})^2}
\end{equation}
and choose $m$ the index for which $|\lambda_m|$ is the biggest of the $|\lambda_k|$'s. Then we take $c_k=1$ for $k\not=m$. We explain now the algorithm that we use to choose a plus or minus sign for $\lambda_k$ as defined in (\ref{lambdak}), with the aim of being able to determine $c_m$ such that (\ref{conditionequation}) is satisfied. We start by choosing a plus sign for $\lambda_m$. Then we choose a minus sign in (\ref{lambdak}) for the $\lambda_k$ for which $|\lambda_k|$ is the second biggest; let us call the index of this term $m_2$. This means that $0\le\lambda_m+\lambda_{m_2}$. For the $\lambda_k$ for which $|\lambda_k|$ is the third biggest -- let us call the index of this term $m_3$ -- we choose a minus sign in case $0\le\lambda_m+\lambda_{m_2}+\lambda_{m_3}$, and otherwise we choose a plus sign, and in this case we have $0\le\lambda_m+\lambda_{m_2}+\lambda_{m_3}$. We continue this way of choosing, always considering the next biggest $|\lambda_k|$, and hence arrive at a global choice of signs for all of the $\lambda_k$, such that $0\le\lambda_m+\sum_{k\not=m}\lambda_k$. Then we determine $c_m$ such that (\ref{conditionequation}) is satisfied, or more specifically such that
\begin{equation} \label{cmequation}
c_m=\sqrt{{(-\sum_{k\not=m}\lambda_k)^2+(\mu(A\ {\rm or}\ B)_m-{\mu(A)_m+\mu(B)_m \over 2})^2 \over \mu(A)_m\mu(B)_m}}
\end{equation}
We choose the sign for $\phi_k$ as defined in (\ref{cosequation}) equal to the sign of $\lambda_k$. The result of the specific solution that we have constructed is that we can take $M_k({\cal H})$ to be rays of dimension 1 for $k\not=m$, and $M_m({\cal H})$ to be a plane. This means that we can make our solution still more explicit. Indeed, we take ${\cal H}=\compl^{25}$ the canonical 25 dimensional complex Hilbert space, and make the following choices
\begin{eqnarray} \label{vectorA}
|A\rangle&=&(\sqrt{\mu(A)_1},\ldots,\sqrt{\mu(A)_m},\ldots,\sqrt{\mu(A)_{24}},0) \\ \label{vectorB}
|B\rangle&=&(e^{i\beta_1}\sqrt{\mu(B)_1},\ldots,c_me^{i\beta_m}\sqrt{\mu(B)_m},\ldots,e^{i\beta_{24}}\sqrt{\mu(B)_{24}},\sqrt{\mu(B)_m(1-c_m^2)}) \\ 
\label{anglebetan}
\beta_m&=&\arccos({2\mu(A\ {\rm or}\ B)_m-\mu(A)_m-\mu(B)_m \over 2c_m\sqrt{\mu(A)_m\mu(B)_m}}) \\
\label{anglebetak}
\beta_k&=&\pm\arccos({2\mu(A\ {\rm or}\ B)_k-\mu(A)_k-\mu(B)_k \over 2\sqrt{\mu(A)_k\mu(B)_k}})
\end{eqnarray}
where the plus or minus sign in (\ref{anglebetak}) is chosen following the algorithm we introduced for choosing the plus and minus sign for $\lambda_k$ in (\ref{lambdak}). Let us construct this quantum model for the data given in Table 1, hence the data collected in \cite{hampton1988}. The exemplar which gives rise to the biggest value of $|\lambda_k|$ is {\it Tomato}, and hence we choose a plus sign and get $
\lambda_{19}=0.0768$. The exemplar giving rise to the second biggest value of $\lambda_k$ is {\it Pumpkin}, and hence we choose a minus sign, and get $\lambda_{20}=-0.0733$. Next comes {\it Yam}, and since $\lambda_{19}+\lambda_{20}-0.0615<0$, we choose a plus sign for $\lambda_{18}$. Next is {\it Green Pepper}, and we look at $0\le\lambda_{19}+\lambda_{20}+\lambda_{18}-0.0503$, which means that we can choose a minus sign for $\lambda_{17}$. The fifth exemplar in the row is {\it Apple}. We have $\lambda_{19}+\lambda_{20}+\lambda_{18}+\lambda_{17}-0.0428<0$, which means that we need to choose a plus sign for $\lambda_8$. Next comes {\it Broccoli} and verifying shows that we can choose a minus sign for $\lambda_{21}$. We determine in an analogous way the signs for the exemplars {\it Raisin}, plus sign, {\it Elderberry}, minus sign, {\it Olive}, plus sign, {\it Peanut}, minus sign, {\it Chili Pepper}, minus sign, {\it Coconut}, plus sign, {\it Watercress}, minus sign, {\it Lentils}, plus sign, {\it Rice}, minus sign, {\it Almond}, plus sign, {\it Acorn}, minus sign, {\it Black Pepper}, plus sign, {\it Mustard}, minus sign, {\it Wheat}, plus sign, {\it Parsley}, minus sign, {\it Root Ginger}, plus sign, {\it Garlic}, minus sign, and finally {\it Mushroom}, plus sign. In Table 1 we give the values of $\lambda_k$ calculated following this algorithm, and from (\ref{cmequation}) it follows that $c_{19}=0.7997$.

Making use of (\ref{vectorA}), (\ref{vectorB}), (\ref{anglebetak}) and (\ref{anglebetan}), and the values of the angles given in Table 1, we put forward the following explicit representation of the vectors $|A\rangle$ and $|B\rangle$ in $\compl^{25}$ representing concepts {\it Fruits} and {\it Vegetables}
\begin{eqnarray}
|A\rangle&=&(0.1895, 0.2061, 0.1929, 0.2421, 0.2748, 0.3204, 0.3373, 0.3441, 0.1222, 0.1165, 0.1252, 0.1291, \nonumber \\
&&0.1002, 0.1182, 0.1059, 0.0974, 0.1800, 0.2308, 0.2967, 0.2823, 0.1194, 0.1181, 0.1245, 0.1128, 0) \\
|B\rangle&=&(0.1154e^{i83.8854^\circ}, 0.1040e^{-i94.5520^\circ}, 0.1484e^{-i95.3620^\circ}, 0.1640e^{i91.8715^\circ}, 0.1120e^{i57.9533^\circ}, \nonumber \\
&&0.1302e^{i95.8648^\circ}, 0.1302e^{-i113.2431^\circ}, 0.1246e^{i87.6039^\circ}, 0.1580e^{-i105.9806^\circ}, 0.1596e^{i99.3810^\circ}, \nonumber \\
&&0.1798e^{i50.0889^\circ}, 0.2112e^{-i86.4374^\circ}, 0.1734e^{-i57.6399^\circ}, 0.2334e^{i18.6744^\circ}, 0.2565e^{-i69.0705^\circ},  \nonumber \\
&&0.2670e^{i104.7126^\circ}, 0.2806e^{-i95.6518^\circ}, 0.2690e^{i98.0833^\circ}, 0.2606e^{i100.7557^\circ}, 0.2670e^{-i103.4804^\circ},  \nonumber \\
&&0.3584e^{-i99.6048^\circ}, 0.2031e^{-i96.6635^\circ}, 0.1630e^{-i61.1698^\circ}, 0.1716e^{i86.6308^\circ}, 0.1565).
\end{eqnarray}
This proves that we can model the data of \cite{hampton1988} by means of a quantum mechanical model, and such that the values of $\mu(A\ {\rm or}\ B)_k$ are determined from the values of $\mu(A)_k$ and $\mu(B)_k$ as a consequence of quantum interference effects. For each $k$ the value of $\theta_k$ in Table 1 gives the quantum interference phase of the exemplar number $k$.

\subsection{Graphics of the Interference Patterns}

In \cite{aerts2009b} we worked out a way to `chart' the quantum interference patterns of the two concepts when combined into conjunction or disjunction. Since it helps our further analysis in the present article, we put forward this `chart' for the case of the concepts {\it Fruits} and {\it Vegetables} and their disjunction `{\it Fruits or Vegetables}'. More specifically, we represent the concepts {\it Fruits}, {\it Vegetables} and `{\it Fruits or Vegetables}' by complex valued wave functions of two real variables $\psi_A(x,y), \psi_B(x,y)$ and $\psi_{A{\rm or}B}(x,y)$. We choose $\psi_A(x,y)$ and $\psi_B(x,y)$ such that the real part for both wave functions is a Gaussian in two dimensions, which is always possible since we have to fit in only 24 values, namely the values of $\psi_A$ and $\psi_B$ for each of the exemplars of Table 1. The squares of these Gaussians are graphically represented in Figures 1 and 2, and the different exemplars of Table 1 are located in spots such that the Gaussian distributions $|\psi_A(x,y)|^2$ and $|\psi_B(x,y)|^2$ properly model the probabilities $\mu(A)_k$ and $\mu(B)_k$ in Table 1 for each one of the exemplars.

For example, for {\it Fruits} represented in Figure 2, {\it Apple} is located in the center of the Gaussian, since {\it Apple} was most frequently chosen by the test subjects when asked {\it Question A}. {\it Elderberry} was the second most frequently chosen, and hence closest to the top of the Gaussian in Figure 2. Then come {\it Raisin}, {\it Tomato} and {\it Pumpkin}, and so on, with {\it Garlic} and {\it Lentils} as the least chosen `good examples' of {\it Fruits}.
\begin{figure}[H]
\centerline {\includegraphics[width=11cm]{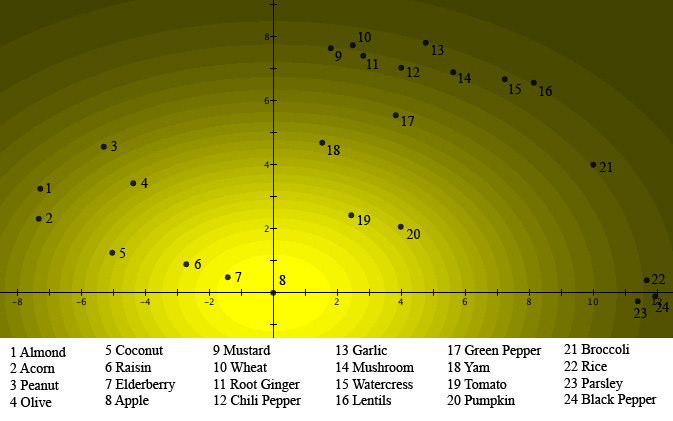}}
\caption{The probabilities $\mu(A)_k$ of a person choosing the exemplar $k$ as a `good example' of {\it Fruits} are fitted into a two-dimensional quantum wave function $\psi_A(x,y)$. The numbers are placed at the locations of the different exemplars with respect to the Gaussian probability distribution $|\psi_A(x,y)|^2$. This can be seen as a light source shining through a hole centered on the origin, and regions where the different exemplars are located. The brightness of the light source in a specific region corresponds to the probability that this exemplar will be chosen as a `good example' of {\it Fruits}.}
\end{figure}
\noindent
For {\it Vegetables}, represented in Figure 3, {\it Broccoli} is located in the center of the Gaussian, since {\it Broccoli} was the exemplar most frequently chosen by the test subjects when asked {\it Question B}. {\it Green Pepper} was the second most frequently chosen, and hence closest to the top of the Gaussian in Figure 3. Then come {\it Yam}, {\it Lentils} and {\it Pumpkin}, and so on, with {\it Coconut} and {\it Acorn} as the least chosen `good examples' of {\it Vegetables}. Metaphorically, we could regard the graphical representations of Figures 2 and 3 as the projections of two light sources each shining through one of two holes in a plate and spreading out their light intensity following a Gaussian distribution when projected on a screen behind the holes. The center of the first hole, corresponding to the {\it Fruits} light source, is located where exemplar {\it Apple} is at point $(0, 0)$, indicated by 8 in both Figures. The center of the second hole, corresponding to the {\it Vegetables} light source, is located where exemplar {\it Broccoli} is at point (10,4), indicated by 21 in both Figures.
\begin{figure}[H]
\centerline {\includegraphics[width=11cm]{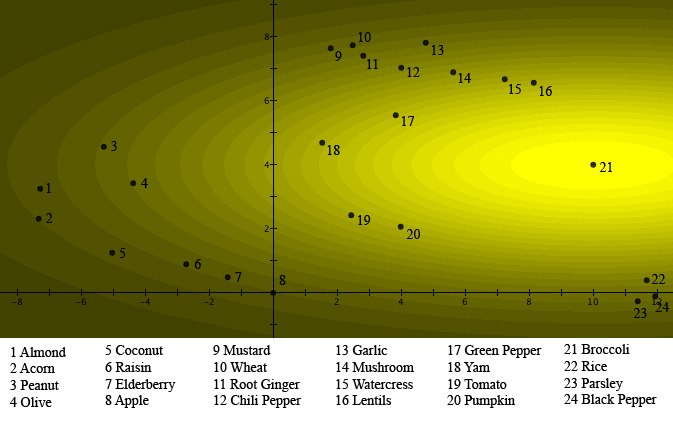}}
\caption{The probabilities $\mu(B)_k$ of a person choosing the exemplar $k$ as an example of {\it Vegetables} are fitted into a two-dimensional quantum wave function $\psi_B(x,y)$. The numbers are placed at the locations of the different exemplars with respect to the probability distribution $|\psi_B(x,y)|^2$. As in Figure 2, it can be seen as a light source shining through a hole centered on point 21, where {\it Broccoli} is located. The brightness of the light source in a specific region corresponds to the probability that this exemplar will be chosen as a `good example' of {\it Vegetables}.}
\end{figure}
\noindent 
\begin{figure}[H]
\centerline {\includegraphics[width=11cm]{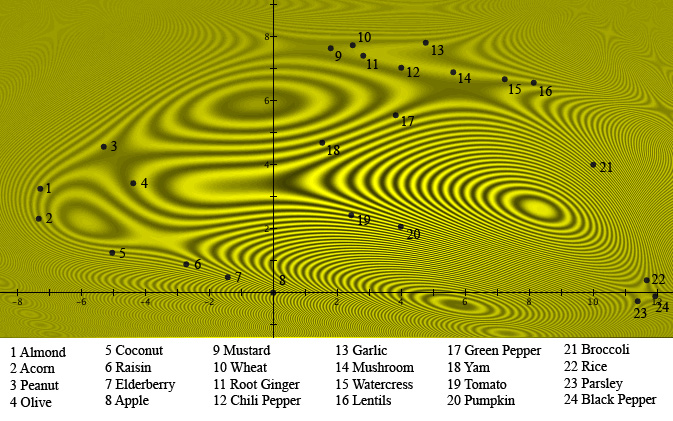}}
\caption{The probabilities $\mu(A\ {\rm or}\ B)_k$ of a person choosing the exemplar $k$ as an example of `{\it Fruits or Vegetables}' are fitted into the two-dimensional quantum wave function ${1 \over \sqrt{2}}(\psi_A(x,y)+\psi_B(x,y))$, which is the normalized superposition of the wave functions in Figures 2 and 3. The numbers are placed at the locations of the different exemplars with respect to the probability distribution ${1 \over 2}|\psi_A(x,y)+\psi_B(x,y)|^2={1 \over 2}(|\psi_A(x,y)|^2+|\psi_B(x,y)|^2)+|\psi_A(x,y)\psi_B(x,y)|\cos\theta(x,y)$, where $\theta(x,y)$ is the quantum phase difference at $(x,y)$. The values of $\theta(x,y)$ are given in Table 1 for the locations of the different exemplars. The interference pattern is clearly visible.}
\end{figure}
\noindent
In Figure 4 the data for `{\it Fruits or Vegetables}' are graphically represented. This is not `just' a normalized sum of the two Gaussians of Figures 2 and 3, since it is the probability distribution corresponding to ${1 \over \sqrt{2}}(\psi_A(x,y)+\psi_B(x,y))$, which is the normalized superposition of the wave functions in Figures 2 and 3. The numbers are placed at the locations of the different exemplars with respect to the probability distribution ${1 \over 2}|\psi_A(x,y)+\psi_B(x,y)|^2={1 \over 2}(|\psi_A(x,y)|^2+|\psi_B(x,y)|^2)+|\psi_A(x,y)\psi_B(x,y)|\cos\theta(x,y)$, where $|\psi_A(x,y)\psi_B(x,y)|\cos\theta(x,y)$ is the interference term and $\theta(x,y)$ the quantum phase difference at $(x,y)$. The values of $\theta(x,y)$ are given in Table 1 for the locations of the different exemplars.
\begin{figure}[H]
\centerline {\includegraphics[width=11cm]{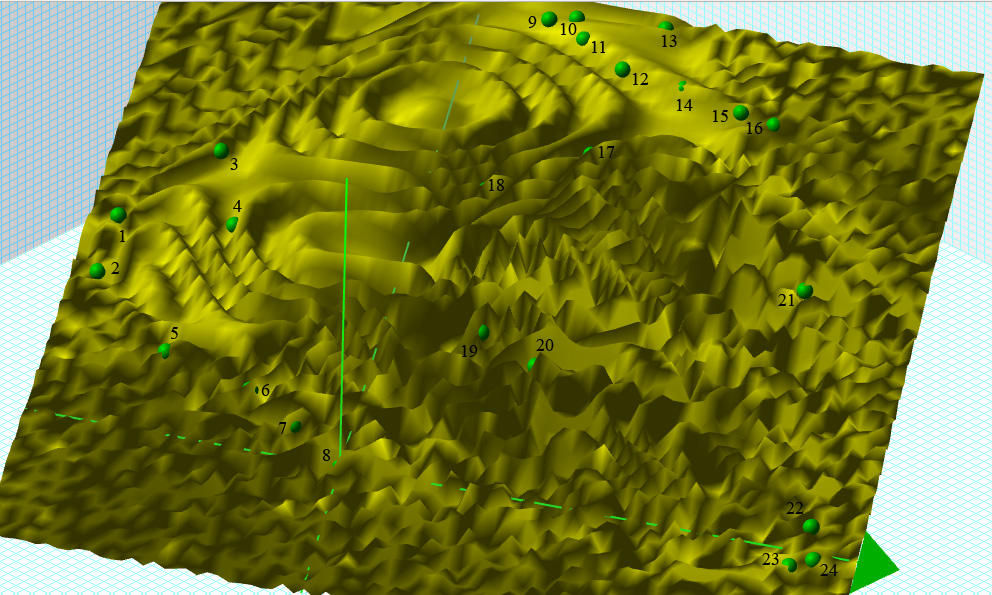}}
\caption{A three-dimensional representation of the interference landscape of the concept `{\it Fruits {\rm or} Vegetables}' as shown in Figure 4. Exemplars are represented by little green balls, and the numbers refer to the numbering of the exemplars in Table 1 and in Figures 2, 3 and 4.}
\end{figure}
\begin{figure}[H]
\centerline {\includegraphics[width=11cm]{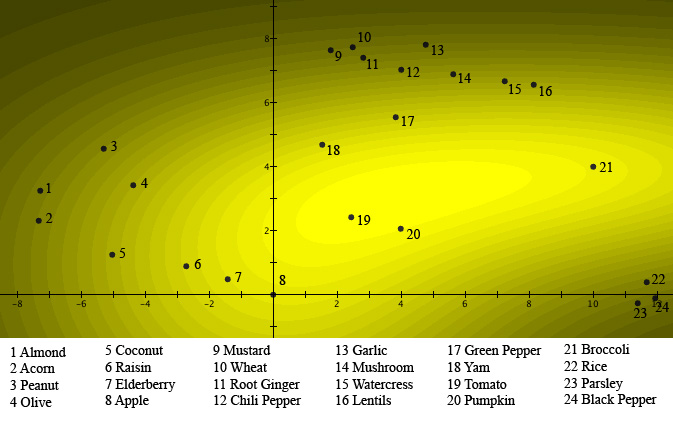}}
\caption{Probabilities $1/2(\mu(A)_k+\mu(B)_k)$, which are the probability averages for {\it Fruits} and {\it Vegetables} shown in Figures 2 and 3. This would be the resulting pattern in case $\theta(x,y)=90^\circ$ for all exemplars. It is called the classical pattern for the situation since it is the pattern that, without interference, results from a situation where classical particles are sent through two slits. These classical values for all exemplars are given in Table 1.}
\end{figure}
\noindent
The interference pattern shown in Figure 4 is very similar to well-known interference patterns of light passing through an elastic material under stress. In our case, it is the interference pattern corresponding to `{\it Fruits or Vegetables}'. Bearing in mind the analogy with the light sources for Figures 2 and 3, in Figure 4 we can see the interference pattern produced when both holes are open. Figure 5 represents a three-dimensional graphic of the interference pattern of Figure 4, and, for the sake of comparison, in Figure 6, we have graphically represented the averages of the probabilities of Figure 2 and 3, i.e. the values measured if there were no interference. For the mathematical details -- the exact form of the wave functions and the explicit calculation of the interference pattern -- and for other examples of conceptual interference, we refer to \cite{aerts2009b}.

\subsection{Explaining Quantum Interference}
The foregoing section shows how the typicality data of two concepts and their disjunction are quantum mechanically modeled such that the quantum effect of interference accounts for the values measured with respect to the disjunction of the concepts. We also showed that it is possible to metaphorically picture the situation such that each of the concepts is represented by light passing through a hole and the disjunction of both concepts corresponds to the situation of the light passing through both holes. This is indeed where interference is best known from in the traditional double-slit situation in optics and quantum physics. If we apply this to our specific example by analogy, we can imagine the cognitive experiment where a subject chooses the most appropriate answer for one of the concepts, for example {\it Fruits}, as follows: `The photon passes with the {\it Fruits} hole open and hits a screen behind the hole in the region where the choice of the person is located'. We can do the same for the cognitive experiment where the subject chooses the most appropriate answer for the concept {\it Vegetables}. This time the photon passes with the {\it Vegetables} hole open and hits the screen in the region where the choice of the person is located. The third situation, corresponding to the choice of the most appropriate answer for the disjunction concept `{\it Fruits or Vegetables}', consists in the photon passing with both the {\it Fruits} hole and the {\it Vegetables} hole open and hitting the screen where the choice of the person is located. This third situation is the situation of interference, viz. the interference between {\it Fruits} and {\it Vegetables}. These three situations are clearly demonstrated in Figures 2, 3 and 4.

In \cite{aerts2007a,aerts2007b,aerts2009c} we analyzed the origin of the interference effects that are produced when concepts are combined, and we provided an explanation that we investigated further in \cite{aertsdhooghe2009}. We will show now that this explanation, in addition to helping to gain a better understanding of the meaning of our basic hypothesis -- that quantum particles behave like conceptual entities -- provides a new and surprising clarification of the interference of quantum entities themselves. To make this clear we need to take a closer look at the experimental data and how they are produced by interference. The exemplars for which the interference is a weakening effect, i.e. where $\mu(A\ {\rm or}\ B) < 1/2(\mu(A)+\mu(B))$ or $90^\circ \le \theta$ or $\theta \le -90^\circ$, are the following (in decreasing order of weakening effect): {\it Elderberry}, {\it Mustard}, {\it Lentils}, {\it Pumpkin}, {\it Tomato}, {\it Broccoli}, {\it Wheat}, {\it Yam}, {\it Rice}, {\it Raisin}, {\it Green Pepper}, {\it Peanut}, {\it Acorn} and {\it Olive}. The exemplars for which interference is a strengthening effect, i.e. where $1/2(\mu(A)+\mu(B)) < \mu(A\ {\rm or}\ B)$ or $\theta < 90^\circ$ or $-90^\circ \le \theta$, are the following (in decreasing order of strengthening effect): {\it Mushroom}, {\it Root Ginger}, {\it Garlic}, {\it Coconut}, {\it Parsley}, {\it Almond}, {\it Chili Pepper}, {\it Black Pepper}, and {\it Apple}. Let us consider the two extreme cases, viz. {\it Elderberry}, for which interference is the most weakening ($\theta=-113.2431^\circ$), and {\it Mushroom}, for which it is the most strengthening ($\theta=18.6744$). For {\it Elderberry}, we have $\mu(A)=0.1138$ and $\mu(B)=0.0170$, which means that test subjects have classified {\it Elderberry} very strongly as {\it Fruits} ({\it Apple} is the most strongly classified {\it Fruits}, but {\it Elderberry} is next and close to it), and quite weakly as {\it Vegetables}. For {\it Mushroom}, we have $\mu(A)=0.0140$ and $\mu(B)=0.0545$, which means that test subjects have weakly classified {\it Mushroom} as {\it Fruits} and moderately as {\it Vegetables}. Let us suppose that $1/2(\mu(A)+\mu(B))$ is the value estimated by test subjects for `{\it Fruits or Vegetables}'. In that case, the estimates for {\it Fruits} and {\it Vegetables} apart would be carried over in a determined way to the estimate for `{\it Fruits or Vegetables}', just by applying this formula. This is indeed what would be the case if the decision process taking place in the human mind worked as if a classical particle passing through the {\it Fruits} hole or through the {\it Vegetables} hole hit the mind and left a spot at the location of one of the exemplars. More concretely, suppose that we ask subjects first to choose which of the questions they want to answer, {\it Question A} or {\it Question B}, and then, after they have made their choice, we ask them to answer this chosen question. This new experiment, which we could also indicate as {\it Question A} or {\it Question B}, would have $1/2(\mu(A)+\mu(B))$ as outcomes for the weight with respect to the different exemplars. In such a situation, it is indeed the mind of each of the subjects that chooses randomly between the {\it Fruits} hole and the {\it Vegetables} hole, subsequently following the chosen hole. There is no influence of one hole on the other, so that no interference is possible. However, in reality the situation is more complicated. When a test subject makes an estimate with respect to `{\it Fruits or Vegetables}', a new concept emerges, namely the concept `{\it Fruits or Vegetables}'. For example, in answering the question whether the exemplar {\it Mushroom} is a good example of `{\it Fruits or Vegetables}', the subject will consider two aspects or contributions. The first is related to the estimation of whether {\it Mushroom} is a good example of {\it Fruits} and to the estimation of whether {\it Mushroom} is a good example of {\it Vegetables}, i.e. to estimates of each of the concepts separately. It is covered by the formula $1/2(\mu(A)+\mu(B))$. The second contribution concerns the test subject's estimate of whether or not {\it Mushroom} belongs to the category of exemplars that cannot readily be classified as {\it Fruits} or {\it Vegetables}. This is the class characterized by the newly emerged concept `{\it Fruits or Vegetables}'. And as we know, {\it Mushroom} is a typical case of an exemplar that is not easy to classify as `{\it Fruits or Vegetables}'. That is why {\it Mushroom}, although only slightly covered by the formula $1/2(\mu(A)+\mu(B))$, has an overall high score as `{\it Fruits or Vegetables}'. The effect of interference allows adding the extra value to $1/2(\mu(A)+\mu(B))$ resulting from the fact that {\it Mushroom} scores well as an exemplar that is not readily classified as `{\it Fruits or Vegetables}'. This explains why {\it Mushroom} receives a strengthening interference effect, which adds to the probability of it being chosen as a good example of `{\it Fruits or Vegetables}'. {\it Elderberry} shows the contrary. Formula $1/2(\mu(A)+\mu(B))$ produces a score that is too high compared to the experimentally tested value of the probability of its being chosen as a good example of `{\it Fruits or Vegetables}'. The interference effect corrects this, subtracting a value from $1/2(\mu(A)+\mu(B))$. This corresponds to the test subjects considering {\it Elderberry} `not at all' to belong to a category of exemplars hard to classify as {\it Fruits} or {\it Vegetables}, but rather the contrary. As a consequence, with respect to the newly emerged concept `{\it Fruits or Vegetables}', the exemplar {\it Elderberry} scores very low, and hence the $1/2(\mu(A)+\mu(B))$ needs to be corrected by subtracting the second contribution, the quantum interference term. A similar explanation of the interference of {\it Fruits} and {\it Vegetables} can be put forward for all the other exemplars. The following is a general presentation of this. `For two concepts $A$ and $B$, with probabilities $\mu(A)$ and $\mu(A)$ for an exemplar to be chosen as a good example of $A$ and $B$, respectively, the interference effect allows taking into account the specific probability contribution for this exemplar to be chosen as a good exemplar of the newly emerged concept `$A\ {\rm or}\ B$', adding or subtracting to the value $1/2(\mu(A)+\mu(B))$, which is the average of $\mu(A)$ and $\mu(B)$.'

The foregoing analysis shows that there is a very straightforward and transparent explanation for the interference effect of concepts. In the following, we will show that this explanation leads to a new understanding of the interference of quantum entities themselves. Also, a detailed analysis of this explanation concerning the double-slit situation for quantum entities allows to understand better our basic hypothesis, namely that quantum particles behave like conceptual entities. Let us consider a typical double-slit situation in quantum mechanics. Figure 7 below presents the interference patterns obtained with both holes open (`$A$ and $B$ open quantum') and only one hole open (`$A$ open $B$ closed' and `$B$ open $A$ closed'), respectively. Rather than present an image of a quantum entity passing through either or both slits `as an object would', we put forward a very different idea, namely the idea that the quantum entity passing through either or both slits `is' the conceptual entity standing for one of these situations. More concretely, we have a quantum entity, let us say `a photon'. This `is' a conceptual entity, hence `the photon is a photon as a concept'. This concept-photon can be in different states, and we will consider three of them: `State $A$ of the concept photon' is the conceptual combination: `the photon passes through hole $A$'.
`State $B$ of the concept photon' is the conceptual combination: `the photon passes through hole $B$'.
`State $A\ {\rm or}\ B$ of the concept photon' is the conceptual combination: `the photon passes through hole $A$ or passes through hole $B$'

To recognize the analogy with our {\it Fruits} and {\it Vegetables} example, we need to consider how {\it Fruits} and {\it Vegetables} are two possible states of the concept {\it Food}. In this analogy, the conceptual combination `the photon passes through hole $A$' corresponds to the conceptual combination `this food item is a fruit', and the conceptual combination `the photon passes through hole $B$' corresponds to the conceptual combination `this food item is a vegetable'. The conceptual combination `the photon passes through hole $A$ or passes through hole $B$' corresponds to the conceptual combination `this food item is a fruit or is a vegetable'.
\begin{figure}[H]
\centerline {\includegraphics[width=14cm]{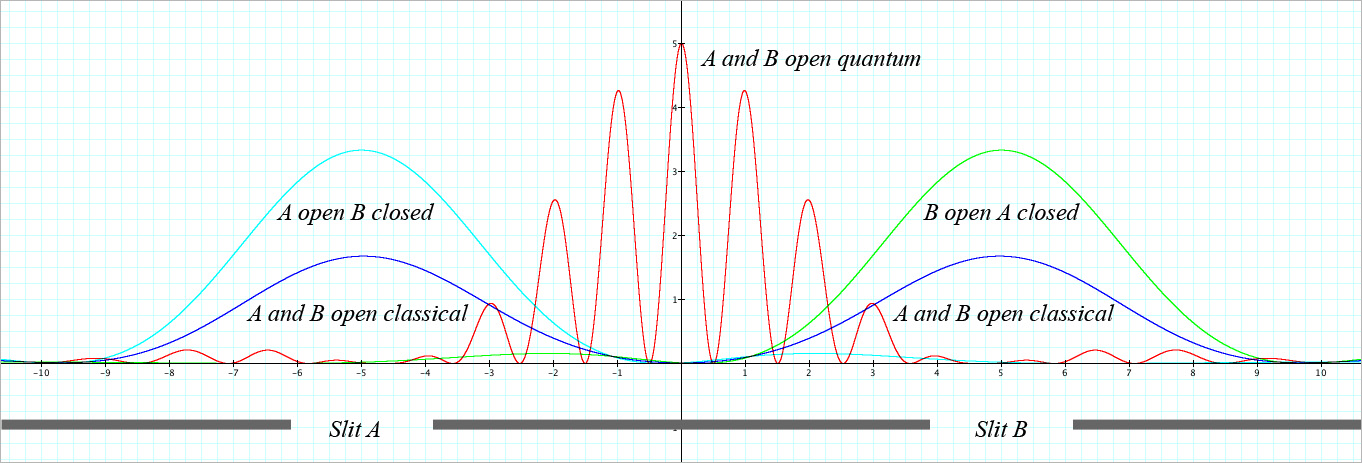}}
\caption{A typical interference pattern of a quantum two-slit situation with slits $A$ and $B$. The `{\it A open B closed}' curve represents the probability of detection of the quantum entity in case only {\it Slit A} is open; the `{\it B open A closed}' curve reflects the situation where only {\it Slit B} is open; and the `{\it A and B open classical}' curve is the average of both. The `{\it A and B open quantum}' curve represents the probability of detection of the quantum entity if both slits are open.}
\end{figure}
\noindent
The photon detected in a spot $X$ on the screen behind the holes is again a specific state of the concept-photon, corresponding to the conceptual combination `the photon is detected in spot $X$'. Compare this to how the different exemplars of Table 1 determine also states of {\it Food}, and hence also states of {\it Fruits} and states of {\it Vegetables} `as concepts'. `Being detected in spot $X$' now corresponds with `spot $X$ being a good example'. Hence, instead of saying that `the photon passing through hole $A$ is detected in spot $X$', we should say `the photon in spot $X$ is a good example of the photon passing through hole $A$'. If we look at the typical interference pattern in Figure 7, we see that on the screen behind slits $A$ and $B$ we have almost zero probability for a photon to be detected in case both slits are open, while we have a very high probability for a photon to be detected on the screen in the center between both slits, completely contrary to what one would expect if photons were objects flying through the slits and subsequently hitting the screen. Let us now analyze this experimental result according to our new interpretation. If both slits are open, this means that the photon is in the state of conceptual combination `the photon passes through {\it Slit A} `or' passes through {\it Slit B}'. And indeed, for a photon hitting the screen in a spot exactly in between both slits, this would be the type of `state of the photon' raising most doubts as to whether it passed through {\it Slit A} or {\it Slit B}. By contrast, photons appearing in the regions behind the slits -- in case both slits are open -- would not make us doubt as to the slit through which they have passed. On the contrary, we can be quite certain that photons showing behind a slit have come through that particular slit, so that this `is not a photon raising doubts about whether it has come through the one or through the other slit'. This means that `a photon in a spot $X$ in the center between both slits {\it is a good example} of a photon having passed through {\it Slit A} or having passed through {\it Slit B}', whereas `a photon in a spot $X$ behind one of the slits {\it is not a good example} of a photon having passed through {\it Slit A} or having passed through {\it Slit B}'.

An interference pattern like that of the concepts therefore makes it possible to deviate from the average probabilities. Analogous to the {\it Fruits} and {\it Vegetables} example, this deviation is due to the fact that the `photon passing through {\it Slit A} or passing through {\it Slit B}' is the conceptual disjunction of `the photon passing through {\it Slit A}' and `the photon passing through {\it Slit B}'. This means that it is completely natural for an estimate of the probabilities of choice with respect to `photon captured in spot $X$' for this conceptual disjunction, for example, to yield high results in the center between the two slits on the detection screen, since it is here that we will have `most doubts as to the slit through which the photon has passed'. Conversely, estimates will be low right behind both slits on the detection screen, since it is `these spots that leave little doubt as to the slit through which the photon has passed'.

We have presented the core elements of our explanation of quantum interference of quantum entities based on the new interpretation we introduced. For the sake of clarity, we have simplified our explanation. Indeed, if we look at the Figures representing the interference of {\it Fruits} and {\it Vegetables}, i.e. Figures 4 and 5, we see that the graphs are at least as complicated as the typical interference graphs of photons, such as the one shown in Figure 7. The reason for this is that the conceptual combinations that we have been considering give rise to more complicated effects than the one we analyzed in detail. We refer to \cite{aerts2009b} for detailed analyses of such more complicated effects. 
But there is more, in \cite{aerts2009c} we analyze how also the conceptual combination `{\it Fruits} and {\it Vegetables}', hence the conjunction instead of the disjunction, can be modeled quantum mechanically similarly to the modeling of the disjunction. Now that we have analyzed in detail the  meaning of the quantum modeling of the disjunction of concepts, namely that it makes it possible to describe the effect of the emergence of the new concept `{\it Fruits} or {\it Vegetables}', and how this effect produces a deviation from the classical value which would be the average, we can understand very well that also a combination of concepts such as `{\it Fruits} and {\it Vegetables}' can be modeled by quantum theory in a similar way. In fact, as our analysis in \cite{aertsgabora2005a,aertsgabora2005b} indicates, a combination of the concepts {\it Fruits} and {\it Vegetables} under the influences of a context, usually expressed by concepts, will lead to different states of this combination. It is these different states that in the formalism of quantum mechanics can be described by different superpositions of states of {\it Fruits} and {\it Vegetables}. Let us give some examples to clarify this. The combination {\it This Fruit tastes like a Vegetable} is such a combination, where {\it taste like} is a context. But also {\it Very special Fruits and Vegetables} is such a combination, where {\it Very Special} is a context. All such combinations can be modeled using the quantum formalism as we did in this section, but the superposition state will be different from the one used for the disjunction. Hence, disjunction is only one of the possible combinations described by superposition and interference.

\section{The One and the Many, Identity and Individuality}
There are two more aspects of quantum physics that are just as important as entanglement and interference in making it a theory fundamentally different from classical physics, namely the presence of Heisenberg uncertainty \cite{heisenberg1927} and the role played by identical quantum particles \cite{castellani1998,vanfraassen1995,frenchkrause2006,dieks2008} and related matters.  In this section we will analyze these aspects and their relation to our basic hypothesis. Before proceeding, we should make the following remark. If we compare the behavior of quantum entities to the behavior of human concepts, we do not expect to find an isomorphic structure for both. We believe that human concepts and their interactions are at a very primitive stage of development as compared to quantum entities and their interactions. This means that we expect to find connections that can have deep explanatory power with respect to fundamental aspects of both situations, the one of human concepts and their interactions and the one of quantum particles and their interactions, but we also expect to find a much less crystallized and organized form for the case of human concepts than for the case of quantum particles. It is within this expectation that one needs to interpret the analysis we put forward in this section.

\subsection{Abstract versus Concrete and Heisenberg Uncertainty} \label{abstractconcrete}

There is a simple basis for a Heisenberg type of uncertainty in the foundation of our interpretation and explanatory framework, namely the incompatibility between `abstract' and `concrete'. It is not possible for a concept to be very abstract and to be very concrete at once. The more abstract a concept is, the less concrete it is, and the more concrete it is, the less abstract it is. Let us give an example to explain what we mean. Consider the concept {\it Cat}. Then, {\it Cat} without any specification, is the most abstract form of this concept. However, if we say {\it This Cat Felix}, and we mean `this particular and unique cat named Felix, the one I can touch and caress with my hand', then this is a most concrete form of the concept {\it Cat}. At the beginning of section \ref{violationbellsinequalities}, we explained how we modeled concepts by using the quantum formalism in \cite{aertsgabora2005a,aertsgabora2005b}, and how we introduced the notion of `state of a concept' to do so. Hence, following up on this, both the most abstract form and the most concrete form of the concept {\it Cat} are considered to be states of the concept {\it Cat}. This means that for each concept, there are states corresponding to more abstract forms of the concept and states corresponding to the more concrete forms. 
In the example of section \ref{entanglement}, {\it Animal} is a more abstract form of {\it Cat}, {\it Cow}, {\it Horse} and {\it Squirrel}, and {\it Food} is a more abstract form of {\it Grass}, {\it Meat}, {\it Fish} and {\it Nuts}. For section \ref{interferencesuperposition}, {\it Fruits} and {\it Vegetables} are more concrete forms of {\it Food}, and all of the concepts in Table 1 are more concrete forms of {\it Fruits} and {\it Vegetables} and also of {\it Food}. 

Let us consider the conceptual environment which is the World Wide Web, and suppose that we google for the word {\it Animal}. On August 25, 2009, this returned 319,000,000 hits, which means that, on that day, there were 319,000,000 webpages listing the word {\it Animal} at least once. In the conceptual environment of the World Wide Web, the totality of combinations of concepts contained in each of these webpages constitutes a state of the concept {\it Animal}, where all other concepts in this combination are conceptual contexts that change the most abstract state of {\it Animal} to the most concrete state for this specific conceptual environment \cite{aertsgabora2005a}. Indeed, webpages containing the word {\it Animal} are the most concrete states of {\it Animal} if we consider the World Wide Web as our specific conceptual environment. Of course, each one of these most concrete states of {\it Animal} is also a most concrete state of many other concepts, namely most of the concepts appearing in the text contained in the relevant webpage. In this sense, if we focus on the conceptual environment which is the World Wide Web, we may consider the collection of all webpages as the analogue for the case of human concepts of what the content of space is for the case of quantum particles. More concretely, if one of the webpages is chosen, opened on a computer screen, and looked at by a person, this is the analogue for the case of human concepts of what a snapshot of space and its content, hence localized states of different quantum entities looked at by an observer, is for the case of quantum particles. The current level of order and structure of the collection of webpages of the World Wide Web is far from that of the collection of quantum particles structured in entities of ordinary matter or in fields of bosonic nature, available to appear as a snapshot of localized states in space. But on a fundamental level, the similarity can be identified. In these snapshots of localized states the type of correlation violating Bell's inequalities can be measured. This shows that both the collection of webpages of the World Wide Web in the case of human concepts and the collections of snapshots of the content of space in the case of quantum particles contain a deep structure that readily allows modeling by the mathematical formalism of quantum theory, based on `coherence' in the case of quantum particles, and based on `meaning' in the case of human concepts. Indeed, on every page of the World Wide Web, each sentence, i.e. each combination of concepts, originates from a source which is the worldview of the individual -- or group of individuals -- who wrote that sentence. Such a worldview is partly personal, but also to a great extent global and resonant with the worldviews of all people. The meaning present in these individual worldviews and in the global worldview is the deep source of the structure of the text on each of the pages of the World Wide Web. In a similar way, it is the nature of the physical coherence between all quantum particles and compositions of such quantum particles that is the deep source of the structure in the set of localized states that can be found in each of the possible snapshots of the content of space experienced by a human observer.
Hence, we should be careful not to make the mistake of thinking that the collection of webpages themselves of the World Wide Web is reality, and by analogy that the collection of snapshots of space presence, i.e. localized states of entities, `is' reality. These snapshots and webpages are only the `localized forms of a reality of coherence' in the case of quantum particles and the `memorized forms of a reality of meaning' in the case of human concepts.

In quantum theory, a localized state of a quantum particle is complementary to a momentum state, i.e. a state where the momentum of the particle is localized in momentum space, and the Heisenberg uncertainty principle stands for the incompatibility of both types of state, i.e. for a quantum particle there are no states that are strongly localized in position space and strongly localized in momentum space. Can we identify the analogue of momentum space for human concepts and, related to this, the Heisenberg uncertainty principle for human concepts? Yes, we can. The more abstract the form of a concept the more it is incompatible with a very concrete form of the same concept. In the foregoing we put forward the collection of all webpages of the World Wide Web as the analogue for the content of position space, relating the choice of one of these pages to a snapshot of the content of position space. In an analogous way, we can make the collection of all abstract forms of human concepts, for example the collection of words in a dictionary, correspond with the snapshots of momentum space and its content. These abstract forms of concepts are the analogue of quantum particles with well-determined momentum, but almost completely non-localized in position space. Let us take a concrete situation to make this clear. This time we consider the conceptual environment of human memories. The most concrete state of a concept then is the state it has in a specific human memory, where the context is defined by all human memory. If two persons communicate with each other by means of the spoken word, then strings of abstract forms of concepts are sent from one human memory to another human memory, triggering these concepts stored in memory, changing their states, or exciting them. The resulting dynamics is what we refer to as communication between two human minds. When quantum particles emitted by a radiating piece of ordinary matter hit another piece of ordinary matter, atoms or molecules in this piece of matter get excited and, when de-exciting, will send out again quantum particles that can eventually be captured by the original piece of ordinary matter. This is a typical situation of matter interacting with quantum particles, and hence also matter interacting through quantum particles with other matter, or matter communicating with matter.

While working, in 2002, on the quantum modeling of concepts \cite{aertsgabora2005a,aertsgabora2005b,gaboraaerts2002}, we noted that this duality -- or complementarity, if we call it by its quantum physics name -- for the case where we consider the World Wide Web as a conceptual environment, or more generally a collection of documents, had been exploited in fields such as `information retrieval' and `semantic analysis' \cite{aertsczachor2004}. Indeed, successful approaches to information retrieval and semantic analysis often consider `documents and terms' as basic ingredients, concentrating on what they call a `document-term matrix' which contains as entries the number of times that a specific term appears in a specific document. Suppose that we label the rows of the matrix by the documents and the columns by the terms, then each row of the matrix can be seen as a vector representing the corresponding document, and each column as a vector representing the corresponding term. If vectors are normalized, the scalar product amongst such normalized vectors is a measure of the similarity of the corresponding documents and terms, and it is also used as such in theories of information retrieval and semantic analysis. In the vector space of vectors representing terms, the documents are represented by the canonical base vectors of this vector space. This means that also the similarity between terms and documents can be calculated by means of the scalar product of the corresponding vectors, and in this way documents can be compared with search terms, and the most relevant documents can be taken to be the most similar ones. This is more or less how today's search engines on the World Wide Web work, although in practice there are many variations on this basic approach. Vector space models for semantic analysis and information retrieval were first introduced by Salton, Wong and Yang \cite{saltonwongyang1975}. Recent examples of such approaches are Latent Semantic Analysis (LSA) \cite{deerwesterdumaisfurnaslandauerharshman1990}, Hyperspace Analogue to Language (HAL) \cite{lundburgess1995}, Probabilistic Latent 
Semantic Analysis (pLSA) \cite{hofmann1999}, Latent Dirichlet Allocation \cite{bleingmichaeljordan2003}, or Word Association Space (WAS) \cite{griffithssteyvers2002}, and connections with quantum structures have now been investigated from different perspectives \cite{widdows2003,widdowspeters2003,vanrijsbergen2004,widdows2006,arafatrijsbergen2007,widdows2008,widdows2009}. 

Let us have a closer look at LSA \cite{deerwesterdumaisfurnaslandauerharshman1990} for which we analyzed correspondences with quantum physics in \cite{aertsczachor2004}. LSA explicitly introduces rank lowering of the document-term matrix by considering the singular value decomposition of this matrix and substituting some of the lower singular values by zero. One reason for introducing this rank lowering technique is to render the sparse matrix of very high rank into a less sparse matrix of less high rank, which makes it more easy to manipulate from a mathematical point of view. There is also an effect of de-noisification, since the original document-term matrix is noisy due to the presence of anecdotal instances of terms. However, there is a third and subtle aspect that interests us specifically with respect to our analysis. If some of the lower singular values are substituted by zero, and the approximated document-term matrix is calculated, it can be shown that the places where the original document-term matrix had zeros, because the terms did not appear in the document, will now contain numbers different from zero. This means that the new document-term matrix reveals `latent' connections between documents and terms. Even if a term does not appear in a specific document, but does appear in many documents similar to this document, the matrix will contain a number different from zero for this term and this document, expressing that, although the term does not appear in the document, it is relevant for the document. To date, LSA has proved one of the most powerful semantic analysis formalisms. The procedures are fully automatic and allow to analyze texts by computers without any involvement of human understanding. LSA produced particularly impressive results in experiments with simulation of human performance. LSA-programmed machines were able to pass multiple-choice exams such as a Test of English as a Foreign Language (TOEFL) (after training on general English) \cite{landauerdumais1997} or, after learning from an introductory psychology textbook, a final exam for psychology students \cite{landauerfoltzlaham1998}. LSA certainly owes much of its potential to its ability to calculate the similarity between a term and a document without the need for the term to appear in the document. The mathematical technique penetrates the meaning structure which is at the origin of the texts to be found in the documents, which are only snapshots of this meaning structure. Hence, by introducing a non-operational mathematical ingredient, the lowering of dimension by means of singular value decomposition and dropping of lower singular values, the LSA approach manages to introduce a mathematical description that is a better model of the underlying meaning structure.

\subsection{The One and the Many} \label{oneandmany}
In the foregoing section we showed how the abstract versus concrete relation of a concept can be understood as the analogue for the case of human concepts of the well-known Heisenberg uncertainty for quantum particles. This analysis also sheds new light on our modeling, in a complex Hilbert space, of the situation of {\it Fruits}, {\it Vegetables} and {\it Fruits or Vegetables} in section \ref{interferencesuperposition}. We have, however, only considered `one concept' as far and different states of this one concept. How does `the many' arise?

In section \ref{entanglement} we dealt with a situation of more than one concept in studying entanglement and the violation of Bell's inequalities. Let us see whether we can understand more clearly the difference between `one' and `many'. If we consider the combination of concepts {\it The Animal eats the Food}, we have at least two entities in mind with respect to this combination, namely an entity related to {\it Animal} and an entity related to {\it Food}. We may even have three entities in mind, i.e. {\it Eat}, if also the different ways of eating -- in other words, the different states of the concept {\it Eat} -- play a role in our interaction with this combination of concepts. If we consider the combination of concepts {\it Fruits or Vegetables}, which we modeled in section \ref{interferencesuperposition}, we mainly have only `one entity' in mind, namely the one corresponding to {\it Food} or to {\it Vegetables}. This is reflected very well in the set of exemplars we chose to consider with respect to the two situations. For {\it The Animal eats the Food} we considered exemplars such as {\it The Cat eats the Grass} and {\it The Squirrel eats the Nuts}, in each of which cases the exemplars also referred to `two entities', one corresponding to {\it Animal} and one to {\it Food}. On the other hand, for {\it Fruits or Vegetables} we considered exemplars such as {\it Coconut}, {\it Broccoli} and {\it Elderberry}, i.e. the exemplars listed in Table 1, and with all of these corresponds `one entity'. When we reflected on the modeling of concepts by means of quantum formalism around 2004, this was one of the hard points to understand. In this period of time, we also struggled with the experimental data we wanted to model, more specifically the data of \cite{hampton1988}, until we finally started to see things more clearly. Let us explain this by means of an example. To point out different aspects, we will consider a combination of concepts that is slightly more complex, namely the following {\it The Animal which is a Herbivore Eats the Tasty Food}. This combination involves five concepts, namely {\it Animal}, {\it Herbivore}, {\it Eats}, {\it Tasty}, and {\it Food}. All of these five concepts play a certain role in how this combination of concepts will affect the typicality of exemplars and application value of features related to it. {\it Herbivore} directly affects {\it Animal}, indeed, certain exemplars of {\it Animal} will have their typicality increased and other decreased by {\it Herbivore}, and certain features of {\it Animal} will have their application value increased and other decreased by {\it Herbivore}, but it will also -- albeit less directly -- affect {\it Food}, in the sense that also certain exemplars of {\it Food} will have their typicality increased and other decreased by {\it Herbivore}, and certain features of {\it Food} will have their application value increased and other decreased by {\it Herbivore}. {\it Tasty} affects {\it Food} directly and {\it Animal}  indirectly, and it also affects {\it Eats}. Indeed, if the food is tasty, we can imagine that the animal eats it in a different way than if it is not. In \cite{aertsgabora2005a}, we explained how all these influences of the different concepts are described by changes in their state, and it is these changes of state that can be modeled by the quantum formalism \cite{aertsgabora2005b}. One new insight, used in \cite{aerts2009c,aerts2007a,aerts2007b}, was  that we could fit the experimental data much better by introducing Fock space, which is the mathematical space used in quantum field theory \cite{bogoliubovlogunovoksaktodorov1990,weinberg1995}. Fock space ${\cal F}$ is defined as a direct sum of Hilbert spaces, where each one of these Hilbert spaces is a multiple tensor product of one Hilbert space. Hence a mathematical expression for Fock space is 
\begin{equation}
{\cal F}=\oplus_{n=1}^m(\otimes_{k=1}^n{\cal H}_k)
\end{equation}
where ${\cal H}_k$ are Hilbert spaces that are all isomorphic to one Hilbert space ${\cal H}$. An element of Fock space, which represents a state of the quantum field, is a superposition of elements of the tensor product components, which means that it is a superposition of states of different numbers of particles, since each vector of $\otimes_{k=1}^n{\cal H}_k$ models a state of $n$ quantum particles.

The introduction of Fock space also greatly enlightened us on the situation of the `one and the many' \cite{aerts2007a,aerts2007b}. Consider again the combination {\it The Animal which is a Herbivore Eats the Tasty Food}, which consists of five concepts. Depending on the situation, different numbers of entities can be connected to this combination of concepts. Fock space, for this combination of concepts consisting of the direct sum of one to five-times tensor products of a Hilbert space $\oplus_{n=1}^{5}(\otimes_{k=1}^n{\cal H}_k)$, is the space which models this in a natural way. Suppose that $|\psi\rangle \in \oplus_{n=1}^{5}(\otimes_{k=1}^n{\cal H}_k)$ describes the state of the combination of concepts {\it The Animal which is a Herbivore Eats the Tasty Food}. This means that
\begin{equation}
|\psi\rangle=\sum_{n=1}^{5}a_ne^{\alpha_n}\otimes_{k=1}^n|\psi_k^n\rangle
\end{equation}
where each of the $|\psi_k^n\rangle \in {\cal H}$ is the state of one of the concepts of the combination, and $a_ne^{i\alpha_n}$ is the amplitude corresponding in Fock space to this state, which means that $a_n^2$ is the weight and $\alpha_n$ is the phase.
If a specific number of entities is primarily involved in a certain conceptual combination, then the biggest weight in Fock space will be attributed to the component of the direct sum containing this number of tensor products of the Hilbert space ${\cal H}$. For the example that we considered, an obvious candidate is the component consisting of a two-times tensor products, referring to the two entities {\it Animal} and {\it Food}. This `two-entity component' of the combination of concepts {\it The Animal which is a Herbivore Eats the Tasty Food} is described in the `two-tensor product component' of Fock space ${\cal H}\otimes{\cal H}$, by the vector $|\psi_1^2\rangle\otimes|\psi_2^2\rangle$, where $|\psi_1^2\rangle$ represents the state of {\it Animal which is a Herbivore} and $|\psi_2^2\rangle$ the state of {\it Tasty Food}, and the weight of this two-entity component is given by $|a_2|^2$, while its phase is given by $\alpha_2$. If, however, the `way of eating' plays a fundamental role, a three-times tensor product, considering the three entities {\it Animal which is a Herbivore}, {\it Eat} and {\it Tasty Food}, might be more appropriate in receiving the biggest weight.

Hence, the general situation is a situation where different numbers of entities in superposition are connected to a specific conceptual combination. The mathematical structure of Fock space as the space for the states of the conceptual combination is well suited for such a modeling \cite{aerts2009a,aerts2007b}. We believe that this is also the basis for understanding the `one' and the `many' with respect to quantum entities. It likewise explains why quantum field theory, where `the number of quantum entities involved' becomes a variable, is a suitable model for the behavior of quantum particles. The quantum entity is described by the quantum field, and the weights corresponding to different numbers of quantum particles, i.e. modes of the field, give rise to the appearance of `the many'. In this respect we should remark the following. It is possible to consider any concept for which there is an entity corresponding to this concept as a state of the concept {\it Thing}. Indeed, {\it Animal}, {\it Food}, {\it Fruits}, {\it Vegetables}, {\it Cat}, {\it Cow}, {\it Horse}, {\it Squirrel} and also all concepts of Table 1 are all exemplars of the concept {\it Thing}, and hence can be described as states of the concept {\it Thing}. Unification attempts with respect to quantum particles are exactly aimed at this, i.e. to find ways such that specific quantum particles appear as states of more general quantum particles or as states of their combinations. This means that concepts such as {\it Animal}, {\it Food}, {\it Fruits}, {\it Vegetables}, {\it Cat}, {\it Cow}, {\it Horse}, {\it Squirrel} and all concepts of Table 1 are already `more concrete' than their `mother concept' {\it Thing}. And indeed, we could describe {\it Animal} as a {\it Thing} which has the main features of {\it Animal}, for example, {\it Is a Living Being}, {\it Can Move Around}, \ldots, and equally so for all other concepts for which there is an entity connected to the concept. Or more drastically, we can describe {\it Animal} as {\it Thing which is an Animal}, where {\it Being an Animal} is considered to be a feature of the concept {\it Thing}. Usually there is no entity connected to concepts expressing features, which means that they are not exemplars of {\it Thing}, and hence cannot be described as states of {\it Thing}. They play only the role of context, and in this way change the state of the concepts that are connected to entities, or more generally, change the state of the concept {\it Thing}. In quantum field theory one has introduced the notion of virtual particles for the type of context-force effect. Not only are virtual particles the analogue of features, but also of other conceptual appearances which only briefly play a role in the dynamics of change of state of the concepts that are connected to entities. Examples are aspects of communication that are very contextual and momentaneous, such as intonations, volume of speech, facial expressions, in the case of vocal communication, and the use of specific writing techniques, such accentuation, size of font (just think of the effect of e-mails written in capitals, which seem to be shouting out at you) in the case of written communication. The concepts that are connected to entities are much more stable and durable as compared to these furtive and very contextual effects of communication, so that we can say they are the predecessors of stable quantum particles.

From a mathematical point of view, the most characteristic aspect of the tensor product components of Fock space is the fact that they delineate Hilbert spaces with compatible elements. More specifically, if we model the combination of concepts {\it The Animal which is a Herbivore eats the Tasty Food} in Fock space, and consider the component consisting of the three-tensor product ${\cal H}\otimes{\cal H}\otimes{\cal H}$, where {\it Animal which is Herbivore}, {\it Eats} and {\it Tasty Food} are each of them modeled in one of the Hilbert spaces of this tensor product, then orthogonal projections used to model changes of state of one of the concepts commute with orthogonal projections used to model changes of state of the other concepts. This is how the `many' appears in a stable way. Compatible concretizations of one and the same abstract concept are conceived as entities and because of their mutual compatibility, these compatible concretizations attain stability with respect to dynamical evolution. Concretely, in the combination {\it The Animal which is a Herbivore Eats the Tasty Food}, the concepts {\it Animal which is a Herbivore}, {\it Eats} and {\it Tasty Food} are all three of them concretizations of {\it Thing}, i.e. states of {\it Thing} which are more concrete due to contextual specifications by means of features. And being in a mutual relation of compatibility, they attain stability with respect to dynamical evolution. More concretely, we could write the combination of concepts {\it The Animal which is a Herbivore Eats the Tasty Food} as {\it Thing, which is a herbivore animal, connects to Thing, which is eating, connects to Thing, which is tasty food}, which shows how the combination is mainly a state of three {\it Things} plus context changing the state of each of the {\it Things}. Equally so, a three-particle state of a component ${\cal H}\otimes{\cal H}\otimes{\cal H}$ of the Fock space of a quantum field is a localization of one-quantum particle as a conceptual entity described by the quantum field, into three more localized states of this one-quantum particle conceptual entity, and the localization and compatibility give stability with respect to dynamical evolution. However, we should take care not to regard such a `three-particle quantum state' as a state that describes three individual objects. Just like {\it Animal which is a Herbivore}, {\it Eat} and {\it Tasty Food} are not three objects -- they are `more concrete' than {\it Thing}, but not objects -- we should not consider a three-quantum particle state to be a state representing three objects. The one-quantum particle conceptual entity described by the three-quantum particle state is `more localized' and `split up into three compatible components', but these three appearances are not objects.

\subsection{Identity and Individuality} \label{identityindividuality}

What about `identical quantum particles'? Will the basic hypothesis of our new explanatory framework help us to shed light on the situation of identical quantum particles? We think the answer is yes. Let us consider a situation where we have a number of entities, say eleven, corresponding to one and the same concept, {\it Animal} for instance. In more standard human language, this means we are considering {\it Eleven Animals}. Suppose further that we consider two states of the concept {\it Animal}, for example the state {\it Cat} and the state {\it Dog}. This means that we have the situation where there are {\it Eleven Animals} of which some are {\it Cats} and some are {\it Dogs}. What are the possible states for these eleven animals? Well, obviously we have {\it Eleven Cats} as a possible state, {\it Ten Cats and One Dog} as another possible state, {\it Nine Cats and Two Dogs}, \ldots, {\it One Cat and Ten Dogs} as a possible state, and {\it Eleven Dogs} as a possible state, i.e. twelve different states in all. We have presented these states in Table 2.
\begin{table}[htb]
\footnotesize
\begin{center}
\begin{tabular}{|lllllll|}
\hline 
\multicolumn{1}{|l}{\it Eleven Animals} & \multicolumn{1}{l}{Google} & \multicolumn{1}{l}{Prob} & \multicolumn{1}{l}{Bose} & \multicolumn{1}{l}{Prob} & \multicolumn{1}{l}{Boltzman} & \multicolumn{1}{l|}{Prob} \\
\hline
{\it Eleven Cats} & 7,700 & 0.2927 & 1 & 0.0833 & 1 & 0.0005 \\
{\it Ten Cats and One Dog} & 250 & 0.0095 & 1 & 0.0833 & 11 & 0.0054 \\
{\it Nine Cats and Two Dogs} & 962 & 0.0366 & 1 & 0.0833 & 55 & 0.0269 \\
{\it Eight Cats and Three Dogs} & 880 & 0.0334 & 1 & 0.0833 & 165 & 0.0806 \\
{\it Seven Cats and Four Dogs} & 3,740 & 0.1422 & 1 & 0.0833 & 330 & 0.1611 \\
{\it Six Cats and Five Dogs} & 411 & 0.0156 & 1 & 0.0833 & 462 & 0.2256 \\
{\it Five Cats and Six Dogs} & 3,310 & 0.1258 & 1 & 0.0833 & 462 & 0.2256 \\ 
{\it Four Cats and Seven Dogs} & 627 & 0.0238 & 1 & 0.0833 & 330 & 0.1611 \\ 
{\it Three Cats and Eight Dogs} & 8 & 0.0003 & 1 & 0.0833 & 165 & 0.0806 \\
{\it Two Cats and Nine Dogs} & 7 & 0.0003 & 1 & 0.0833 & 55 & 0.0269 \\ 
{\it One Cat and Ten Dogs} & 5 & 0.0002 & 1 & 0.0833 & 11 & 0.0054 \\
{\it Eleven Dogs} & 8,410 & 0.3197 & 1 & 0.0833 & 1 & 0.0005 \\ 
\hline
\end{tabular}
\end{center}
\caption{The twelve states that are formed for the concept {\it Animal} appearing as {\it Eleven Animals} with respect to two possible states of {\it Animal}, namely {\it Cat} and {\it Dog}. Google and Prob give the distribution of these states on webpages of the World Wide Web. Bose and Prob are the Bose Einstein distribution of these states, and Boltzman and Prob are the Maxwell Boltzman distribution of these states.}
\end{table}
When we consider {\it Eleven Animals}, each one of these eleven {\it Animals}, as a concept, is completely identical to the other ten {\it Animals}. The same holds for the {\it Cats} and {\it Dogs}. Hence, we see here that our basic hypothesis, namely that quantum entities are conceptual entities, offers a simple and straightforward explanation of the `identity of quantum particles'. It also follows from our basic hypothesis that these twelve states are really `all there is'. Indeed, it is precisely because of the `nature of the identity of concepts' that there is no underlying reality defining more precise states for these concepts.

Is it possible to find out whether a quantum mechanical Bose Einstein statistics or a classical Maxwell Boltzman statistics applies to this situation? We can indeed make a strong case for a quantum Bose Einstein statistics. Let us suppose that {\it Cat} and {\it Dog} are equally probable states. If there was an underlying reality where these states can be further specified described by a Maxwell Boltzman statistics, the number of elements in an $(11-n)\times$cat and $n\times$dog situation, for $n$ being a number between 0 and 11, would then be given by  $11!/n!(11-n)!$. We have calculated these quantities and they can be found in Table 2 in the column under Boltzman. Their total sum equals $2^{11}$, and hence we can calculate the probabilities $11!/(n!(11-n)!\cdot2^{11})$ for each of the situations of eleven animals of Table 2 being realized. These probabilities for the different possible realizations are given in Table 2 in the last column. The biggest probability, 0.2256, corresponds to the cases of {\it Six Cats and Five Dogs} and {\it Five Cats and Six Dogs} and the smallest probability, 0.0005, corresponds to the cases of {\it Eleven Cats} and {\it Eleven Dogs}. However, a Bose Einstein statistical description of the situation attributes equal probability, 0.0833, to each one of the cases, as shown in Table 2 under Bose and Prob. We can also conduct a Google search for each of the possibilities. On September 7, 2009, this produced the following results: for {\it Eleven Cats}, 7,700; for {\it Ten Cats and One Dog}, 250, \ldots; for {\it One Cat and Ten Dogs}, 5, for {\it Eleven Dogs}, 8,410. We have listed all of them in Table 2 under Google, and calculated the corresponding probabilities, which are listed in the second column in Table 2. It is easy to see that the Google probabilities are much closer to the Bose Einstein probabilities than to the Maxwell Boltzman probabilities. In fact, the Google statistics is less regular than the Bose Einstein, but follows essentially a similar structure, i.e. one in which mainly the different cases are equally probable. The exception is that the {\it Eleven Cats} and the {\it Eleven Dogs} are much more probable, but this is certainly partly due to {\it Eleven Cats} and {\it Eleven Dogs} containing only two concepts, while all the other cases contain five concepts. As a consequence, Google counts are higher for these two cases. Furthermore, there are some fluctuations, namely {\it Seven Cats and Five Dogs} and {\it Five Cats and Six Dogs} come out with higher probabilities than the other combinations. We did a similar calculation for {\it Horse} and {\it Cow} instead of {\it Cat} and {\it Dog} and a very comparable pattern showed up. The two extremes {\it Eleven Horses} and {\it Eleven Cows} yield higher probabilities, and all the other results are similar. Again, there are two fluctuations, but not for the same combinations of numbers, which shows that in both cases these deviating cases are fluctuations.

Taking the World Wide Web once again as a possible conceptual environment for human concepts, we can see that within this conceptual environment the identity of concepts leads to a statistics which is of the Bose Einstein type and not of the Maxwell Boltzman type. If we were to count the numbers of cats and dogs that people have living in their homes, we would find a distribution of the Maxwell Boltzman type and not of the Bose Einstein type. The reason is that these cats and dogs are individuals and hence not identical, whereas {\it Dogs} and {\it Cats} as concepts are not individuals and are identical. We believe that the fact that quantum entities behave like completely identical entities, and moreover, exactly in the same way as concepts behave with respect to identity, is a strong argument in favor of the basic hypothesis of our new quantum interpretation and explanatory framework. As far as we know, there is no other interpretation of quantum mechanics which contains an explanation of the specific behavior of identical quantum particles. 

In this respect, we will give another example of how our basic hypothesis sheds light on the behavior of identical quantum particles. More specifically, we will consider the situation of the collision and consequent scattering of quantum particles which many textbooks include as a highly dramatic illustration of the mysterious behavior of identical quantum particles. If quantum particles are made to collide, quantum mechanics predicts that the resulting cross sections of particles scattered due to the collision will be drastically different, depending on whether the quantum particles are identical or distinguishable, and this prediction is confirmed by experiment. For example, for a scattering angle of ${\pi \over 2}$ in the center of mass configuration, the scattering cross section for identical bosons is twice that for distinguishable quantum particles \cite{bromleykuehneralmqvist1961}. Let us show that, in a similar way, identical concepts behave very differently from distinguishable concepts. We turn to our example of section \ref{interferencesuperposition}, and consider the concepts {\it Fruits} and {\it Vegetables} and the exemplars of these concepts in Table 1. However, this time we use the Yahoo search engine to count webpages to make our point. The day of our search is September 12, 2009. We consider different combinations of the exemplars and search the number of webpages containing each of the pairs of exemplars and the two distinguishable concepts {\it Fruits} and {\it Vegetables}. The results are given in the first column of Table 3.
\begin{table}[htb]
\footnotesize
\begin{center}
\begin{tabular}{|lllll|}
\hline 
\multicolumn{1}{|l}{{\it Pairs of Examplers}} & \multicolumn{2}{l}{{\it Fruits, Vegetables}} & \multicolumn{2}{l|}{{\it Fruits or Vegetables}}
 \\
\hline
{\it Yam, Watercress} & 41,400 & 0.0016 & 124 & 0.0011 \\
{\it Mustard, Raisin} & 672,000 & 0.0267 & 734 & 0.0068 \\
{\it Peanut, Broccoli} & 1,120,000 & 0.0445 & 14,600 & 0.1347 \\
{\it Apple, Wheat} & 2,530,000 & 0.1005 & 31,100 & 0.2870 \\
{\it Garlic, Mushroom} & 3,150,000 & 0.1251 & 2,730 & 0.0252 \\
{\it Olive, Green Pepper} & 6,020,000 & 0.2392 & 17,600 & 0.1624 \\
{\it Rice, Tomato} & 5,650,000 & 0.2245 & 34,900 & 0.3221 \\ 
{\it Almond, Black Pepper} & 2,940,000 & 0.1168 & 1,700 & 0.0157 \\ 
{\it Chili Pepper, Parsley} & 1,550,000 & 0.0616 & 1,670 & 0.0154 \\
{\it Pumpkin, Coconut} & 980,000 & 0.0389 & 2,830 & 0.0261 \\ 
{\it Lentils, Root Ginger} & 489,000 & 0.0194 & 329 & 0.0030 \\
{\it Acorn, Elderberry} & 27,600 & 0.0011 & 33 & 0.0003 \\ 
\hline
\end{tabular}
\end{center}
\caption{The pair of concepts {\it Fruits}, {\it Vegetables} as two distinguishable concepts `interact' with the pairs of different exemplars. The number of webpages containing each of the different pairs of exemplars plus {\it Fruits} plus {\it Vegetables} was counted using the Yahoo search engine (column one), and the probabilities were calculated (column 2), under the hypothesis that all webpages were equally likely to be chosen. Columns 3 and 4 contain analogous data and calculations, this time the pair of identical concepts {\it Fruits or Vegetables} and {\it Fruits or Vegetables} `interacting' with each of the pairs of different exemplars.}
\end{table}
In the second column we have calculated the probabilities related to the numbers of the first column, each probability being equal to the number of webpages divided by the total number of webpages for all pairs of exemplars. Column three of Table 3 shows the number of webpages containing each of the pairs of exemplars plus the concept combination {\it Fruits or Vegetables}, and column four shows the corresponding probabilities, again each probability being equal to the number of webpages divided by the total number of webpages for all pairs of exemplars. \begin{figure}[H]
\centerline {\includegraphics[width=7cm]{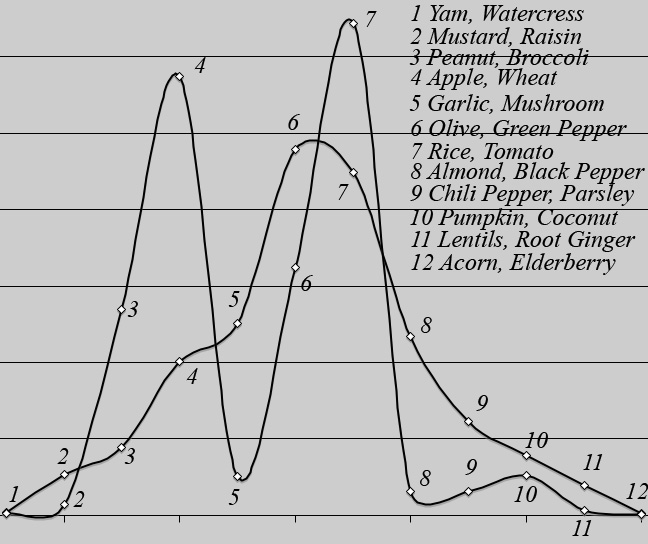}}
\caption{The weights functions for the different pairs of exemplars with respect to the pairs of distinguishable concepts {\it Fruits} and {\it Vegetables}, and the pair of identical concepts {\it Fruits or Vegetables} and {\it Fruits or Vegetables}.}
\end{figure}
\noindent
Figure 8, representing both collections of probabilities, reveals the big difference in behavior in both cases. The biggest weight of the situation of distinguishable concepts {\it Fruits} and {\it Vegetables} goes to the pair {\it Olive, Green Pepper}, while for the situation of identical concepts {\it Fruits or Vegetables} the biggest weight goes to the pair {\it Rice, Tomato}. This means that the pair {\it Olive}, {\it Green Pepper} scores higher in an environment that considers both {\it Fruits} and {\it Vegetables}, as distinguishable concepts, while the pair {\it Rice, Tomato} scores higher in an environment where the question is about {\it Fruits or Vegetables} and where we do not know whether we are dealing with {\it Fruits} or {\it Vegetables}, i.e. an environment where {\it Fruits} and {\it Vegetables} are treated as identical concepts, both being states of the concept {\it Fruits or Vegetables}. Coming back to the analysis we made in section \ref{interferencesuperposition}, we can indeed imagine that {\it Rice, Tomato} are more the types of exemplar that raise doubt as to whether they are {\it Fruits} or whether they are {\it Vegetables}, while this is not the so much the case for {\it Olive, Green Pepper}. Of course, this is only one aspect of the many different and subtle aspects that play a role in how weights are determined based on the frequency of occurrence of these concepts in pages on the World Wide Web. We have looked at several other such situations comparing distinguishable concepts with identical ones, and always found substantial differences between the weights corresponding to the two situations, which indicates that the effect is structural and not coincidental.

We can now put forward a more general scenario for the question of how `the many' starts to get formed out of `the one', and distinguish two types of situation. In the first type of situation, the concepts are `in a dynamic flow of interaction', and in the second type of situation, the concepts are `structured in a memory'. The two types of situation can be distinguished by observing `how identity is handled'. If in the first type of situation identical concepts appear, they can be in the same state. This is explained by the situation we analyzed above for the {\it Twelve Animals}. If we consider the second type of situation, `structured in memory', for example by looking up the word {\it Cat} in a dictionary, we can see that it does not appear in two places. This means that {\it Animal} and {\it Animal} as two identical concepts cannot be in the same state {\it Cat}. Structured in a memory, the state of {\it Animal} which is {\it Cat}, only appears once. Situations structured in memory differ essentially from dynamic flow situations in that memory attributes unique places to states of concepts. Hence, once in memory, two identical concepts cannot be in the same state. This `exclusion principle' is manifestly present in a computer memory, which refuses a name for a file when this name has already been given to another file in the same directory or folder. For quantum entities, the two types of situation are linked to boson and fermion quantum entities. Quantum entities are bosons when they are in states contained in dynamic flow, in quantum language, `when they are carriers of the interaction field', and they are fermions when they are in states contained in ordinary matter. According to our interpretation, ordinary matter is the equivalent of the quantum level of memory. The Pauli exclusion principle for fermions, quantum entities of ordinary matter, expresses that each state of a quantum entity addresses a unique place. It has been shown that Pauli's exclusion principle is fundamental for matter to be stable \cite{dysonlenard1967,lenarddyson1968,lieb1976,lieb1979}, and also that bosonic matter is unstable regardless of the dimension of space \cite{muthapornmanoukian2004}. 

We already mentioned that, with respect to our comparison between human concepts and quantum entities and their respective interfaces, memory structures for human concepts, and ordinary matter for quantum entities, we expect to find a much less advanced organization for the case of human concepts than for the case of quantum particles. If we consider again the Word Wide Web as our example of a memory environment for human concepts, then the webpages of the World Wide Web are the equivalent for the case of human concepts of snapshots of the material and energetic content of space for the case of quantum particles. Each webpage has its unique and individual location somewhere on a server, which is the primitive version of the Pauli exclusion principle for the World Wide Web memory structure related to human concepts. When a human mind interacts with such a webpage and concepts and combination of concepts are exchanged between the memory structure of the human mind and the webpage, in this reality of dynamic flow related to this interaction, the meaning content of the webpage is not bound to a Pauli exclusion principle. For example, there is no problem at all for different human minds to interact with the meaning content of the same webpage. This is already the case when different people are watching the same computer screen at the same time. We can therefore see that the fundamental duality -- bosonic statistics presenting itself in the realm of dynamic flow of human concepts as well as in the realm of dynamic flow of quantum particles, and fermionic statistics governing the realm of interfaces of this dynamics flow, memory structures in the case of human concepts and ordinary matter in the case of quantum particles -- does indeed present itself already on the level of our comparison. However, there has been a strong organizational phase in the case of quantum particles, introducing space as the theatre of the localized states of quantum entities and ordinary matter. Further elaboration of the explanatory framework that we propose here will be necessary to gain a deeper understanding of the underlying principles of these aspects of the organization of quantum particles and their interactions. For example, attempts should be made to understand the spin-statistics theorem \cite{fierz1939,pauli1940}, which shows that quantum particles with integer spin values behave according to a Bose Einstein statistics, while particles with half integer spin values behave according to a Fermi Dirac statistics, in the light of the explanatory framework we propose. 

The original proof of the spin-statistics theorem relies heavily on relativistic quantum field theory and is far from transparent \cite{fierz1939,pauli1940}. Later elaborations have also failed to clarify much of the strongly technical nature of the original evidence \cite{streaterwightman1989,pauli1950,feynman1961,ducksudarshan1998}. However, on a more global level, the correspondence can be analyzed as follows. One of the main differences between quantum particles with half entire value of the spin and quantum particles with entire value of the spin, is that the wave function of the former changes sign under the influence of a $360^\circ$ rotation, while the latter does not change sign under the influence of such a rotation. For two identical quantum particles, the identity of the particles is expressed mathematically by requiring the wave function of the two particles to be invariant for permutation of both particles. This has one of two outcomes, to wit: the wave function is anti-symmetric,  changing sign by permutation, or symmetric, remaining unchanged by permutation. The spin-statistics theorem proves that the `changing sign' and `remaining unchanged' in both cases correspond, for the `sign changing situation for rotation' manifested with particles with half integer spin coincides with the `sign changing situation for permutation' for fermions, and the `remaining unchanged situation for rotation' manifested with integer spin coincides with the `remaining unchanged situation for permutation' for bosons. There have been attempts to prove the spin statistics theorem by relying less on technical aspects of relativistic field theory and more on topological aspects of the situation \cite{feynmanweinberg1987,baezodyrichter1995,hara2003}, without however being able to make the full circle. It has also been shown that the fact that space has at least three dimensions is crucial, and that for a two or one-dimensional space, para-statistics, i.e. mixtures of Fermi and Bose statistics, are possible. Quantum particles existing in such two-dimensional space situations have been called anyons \cite{leinaasmyrheim1977,wilczek1982}, and in recent years an interference situation was realized constituting a direct observation of fractional statistics of anyons \cite{caminozhougoldman2005}.

For the case of human concepts, although we can express the requirement of identity in general terms, the situation of human concepts and their interface of memory structures has not evolved sufficiently to contain a structure where rotational invariance may be expressed in general terms. This is also the reason that no equivalent of spin exists on this level. However, for quantum particles, spin and rotational invariance are the examples of internal symmetries. We can try to identify internal symmetries for the case of human concepts and memory structure, in which case an equivalent of the spin statistics theorem would consist in expressing that the effect of internal symmetry on the Hilbert space used to model the human concepts should coincide with the effect of exchange due to identity. It is an approach to the situation that we intend to investigate in depth in the future.

What about individuality? On several occasions, we have put forward the example of the World Wide Web as a conceptual environment, pointing out that understanding what individuality is would be of help to conduct an in-depth analysis of the overall consequences of our new quantum interpretation and explanatory framework. Having come this far, we can now be more specific on this point. What the example shows us, is that basic questions about the nature of quantum entities -- such as about identity and individuality, but also about localization and objectivity, and about micro and macro, as we will see in section \ref{micromacro} -- should and can be approached with more subtlety than has commonly been the case so far. With respect to localization, we have already been more explicit in the sense that retrieving and showing a webpage on a computer screen for it to be watched, read and meaningfully connected to by a human mind can be seen as playing a similar role for the case of human concepts to the role that the localization happening during a measurement in quantum mechanics plays for the case of quantum particles. Space is an environment where quantum entities exist in localized states and they do so in a much more structured and organized manner than the collection of webpages interacting with human minds. But the questions about identity, individuality, localization, objectivity, micro and macro, as far as fundamental aspects of these questions are concerned, can be put forward also for the case of human concepts in its more primitive organizational situation as compared to quantum particles. Indeed, its primitive organizational situation enhances the chances of being able to identify the fundamental aspects of these questions.

The World Wide Web is estimated currently to contain more than 20 billion webpages \cite{dekunder2009}, with a world population of 6,784 billion people, of whom an estimated number of 1 to 2 billion are connected to the internet. This means that not all webpages can be looked at by a human mind at a specific moment of time, and the discrepancy between the number of webpages and the number of minds potentially interacting with them is bound to increase over time. It also means that a substantial part of the meaning content of the World Wide Web is potential, in the sense that it is potentially available to be connected to a human mind, not all of this potential being actualized at any specific moment of time. This situation is by no means restricted to the World Wide Web. Just think of the numerous articles and books that are not being read at this specific moment, most likely greatly outnumbering those that are being read. Also, the human mind itself contains at any moment much more potential meaning content than what is being actualized. And again, in the case of computers, what is used at the interface at a certain moment is only a fraction of their potential. It is a general property of organized memory systems that much more meaning content is stored in a potential way than is being used in an actualized way. From this perspective, it is not surprising to encounter a similar situation for quantum particles and their interfaces. As we mentioned before, we consider space as a theatre of such specific actualized states of quantum particles, in analogy with a webpage read by a human mind being a specific state of the conceptual environment which is the World Wide Web. It is worth mentioning an aspect of the actual functioning of search engines of the World Wide Web which is different from how snapshots of space give rise to localized states of quantum entities. Search engines look for similarity between the search term and webpages, and put forward a ranking accordingly, which is basically a deterministic approach. Contrary to this, the localization of quantum particles within snapshots of space is not deterministic. There are good reasons to believe that the way human minds interact with each other and with strings of human concepts is also non-deterministic. It is quite possible that the deterministic functioning of actual search engines is a provisional stage of how the World Wide Web is explored. It can indeed be argued that there are definite advantages when it comes to having meaning flourish in a optimal way in such an environment as the World Wide Web to have search engines that make use of non deterministic but weighted ranking, in the sense that the actual ranking should be substituted by a probabilistic ranking \cite{aerts2005}. We will not elaborate on this any further in the present article but  mention it mainly as an introduction to the following line of reasoning.

We have recurrently used concepts combined by the connective `or', such as {\it Fruits or Vegetables}. In \cite{aerts2009c} we showed, more thoroughly than we have been able to do in the present article, the connection between `or' and the use of superposition in Hilbert space. Let us now consider two webpages and call them {\it Webpage A} and {\it Webpage B}. Is it possible to give a meaning to `{\it Webpage A or Webpage B}'? Clearly `{\it Webpage A or Webpage B}' is not a webpage, but it may be used to indicate that, in the realm of potentiality, {\it Webpage A} and {\it Webpage B} can both be actualized, i.e. change their state of availability for the human mind to be read to that of actually being read. If a probabilistically weighted search engine existed, these events would involve probabilities. Since webpages are considered to be the optimally concrete concepts within the conceptual environment which is the World Wide Web, such a conceptual entity as `{\it Website A or Website B}' is not a website, because it is not an optimally concrete conceptual entity.
In the case of {\it Fruits or Vegetables}, we can also consider {\it Fruits or Vegetables} as more abstract than {\it Fruits} and {\it Vegetables} separately. Let us consider an analogous situation for quantum particles. For two localized states $\psi_A(x,y,z)$ and $\psi_B(x,y,z)$ of a quantum particle, the state ${1 \over \sqrt{2}}(\psi_A(x,y,z)+\psi_B(x,y,z))$ is usually not at all a localized state. There is, however, another way of looking at this situation. Suppose that a human mind is in interaction with -- the meaningfully connected subject matter of -- {\it Webpage A} and {\it Webpage B}, reading both webpages, interested in the content of both, and even going from one to the other and back. As a result of this interaction of the human mind with both webpages, the content of both webpages may become mixed in a meaningful manner, so that eventually a new {\it Webpage C} may arise. This {\it Webpage C} can most probably be modeled by superposition of {\it Webpage A} and {\it Webpage B}. This is borne out by \cite{aerts2009c}, where we showed that `superposition' serves to model not only the connective `or' but also the connective `and', for example, as well as more complicated `conceptual combinations of two concepts'. This indicates that, on the level of what happens with the World Wide Web and its continuous dynamic growth, superposition is able to model the creation of new webpages.

In section \ref{abstractconcrete} we explained how `semantic analysis' and `information retrieval' make use of vector space models in a way that is very similar to how quantum mechanics uses the Hilbert space formalism. There is another field of research which is relevant for the connection we explore in the present article, namely `decision theory' as it is used in the field of psychology. We were inspired to apply quantum structures to model a human decision process by the fact that in such a decision process the `possible outcomes of the decision do not exist prior to the decision being taken'. Classical probability is a formalization of a situation where these possible outcomes `do exist' but `we lack knowledge about them'. Quantum probability is exactly the opposite in this respect, for it describes processes where indeterminism appears `not due to a lack of knowledge of an already existing set of events', but `due to the probability connected to potential (and hence not yet existing) events'. This is at the origin of the quantum decision model we proposed in \cite{aertsaerts1994}. Meanwhile, quantum structures have been applied to the modeling of human decision processes in a general way and for several aspects of such decision processes \cite{busemeyerwangtownsend2006}, notably to model the disjunction effect 
\cite{busemeyermatthewwang2006,pothosbusemeyer2009}, conjunction fallacy \cite{franco2009} and the violation of the Savage `sure thing principle' \cite{khrennikovhaven2009}, while we have analyzed how these quantum structures of decision processes relate to the way we have conceptually structured our world, by means of concepts and combinations of concepts \cite{aertsdhooghe2009}.
If we take into account that it is not only the overall structure of concepts and their combinations which entail quantum structure, but also decision processes related to actualizing situations which were potential before being actualized, we can elaborate our explanatory framework for quantum theory accordingly, and propose an interpretation of the so-called `collapse of the wave function' as a process which describes the change of potentiality to actuality, with respect to the interface for which this actuality is defined.
 
Let us now try to understand what individuality means within this framework. In a way, we might say that a freshly created webpage is an individual, although it is definitely a combination of concepts given place in a memory structure. However, for a well-established and well-known webpage like the Wikipedia page on Johann Sebastian Bach \url{http://en.wikipedia.org/wiki/Johann_Sebastian_Bach}, it would be less obvious to attribute the connotation of individual. It is likely to have been copied many times, so that we may assume that there are identical versions in many different places. But is this really so? There is one feature of the World Wide Web that makes it quite unique compared to earlier forms of artificial memory structures, namely it is being updated continuously. This means that, in our example, Bach's webpage may come in all sorts of variations of its originally conceived form. For the human memory, this is certainly the case, since it is also constantly changing, and individuality might well be deeply connected to this aspect. We will not make an attempt to give a complete analysis of what individuality means within the interpretation and explanatory framework we put forward in the present article. What we want to show most of all is that the this framework contains the fine structure that is required to give place to the notion of individuality, although it is not an objective notion. Whether a specific combination of concepts is an individual depends on the details of the conceptual environment in which this conceptual combination is located. This is the case for combinations of human concepts, and we suppose this is also the case for quantum particles and their combinations.

\subsection{Micro and Macro and Schr\"odinger's Cat} \label{micromacro}

One `way of speaking about the explanatory framework we put forward' is by introducing a different relation between micro and macro than the one commonly used, and by characterizing the microscopic realm and the macroscopic realm in a different way than the usual one. The macroscopical realm then consists of entities which are human concepts and whose behavior is also described by the quantum mechanical formalism \cite{aertsgabora2005a,aertsgabora2005b}, being the macroscopic equivalents of quantum entities in the microscopic realm. The macroscopic realm also consists of the memory structures that are sensitive to these human concepts, such as minds, computer memories, but also other forms of memory, such as dictionaries and the World Wide Web collection of webpages.
These would be the macroscopic equivalents of what ordinary matter is in the microscopic realm. The macroscopic realm consists of a third type of entities, namely those that do not participate in the communication dynamics between memory structures, and hence are not sensitive to human concepts. Examples are ordinary matter which is not a macroscopic memory structure, so not only dead matter, but also plants and animals, i.e. all entities that cannot communicate by means of human concepts. We could also restrict this collection to `dead matter' if we enlarge the macroscopic communication scheme to also include communication by any type of semiotic sign, i.e. not only by human concepts. In this case, the macroscopic of what ordinary matter is in the micro-realm are the `interfaces' for this communication, i.e. also animals, plants, and computer interfaces. The microscopic realm consists of entities which are the quantum particles, their macroscopic equivalents being human concepts -- or more generally semiotic signs --, and of the material entities sensitive to interactions with these quantum particles, i.e. `all' material entities that we know, made of ordinary matter. They are the microscopic equivalents of what in the macroscopic realm are the memory structures -- or more generally the communication interfaces for semiotic signs. This means that a material entity, made of ordinary matter, whatever its size, `only' interacts in a quantum dynamical way with quantum particles, since it is the microscopic equivalent of what memory structures, e.g. dictionaries or computer memories, or human minds are in the macroscopic realm.

What about Schr\"odinger's Cat paradox? {\it Cat} is a material entity consisting of ordinary matter, with an animal's mind, but as such it is a member of the microscopic realm. But {\it Cat} is also a human concept, and as such it is a member of the macroscopic realm. And {\it Cat} as a human concept is indeed a superposition of {\it Dead Cat} and {\it Living Cat}. Using Google once again, we can even estimate the superposition coefficients. There were 495,000 pages containing the conceptual combination {\it Dead Cat} and 29,400 pages containing the conceptual combination {\it Living Cat} on August 1, 2009. In a Hilbert space ${\cal H}$, describing {\it Dead Cat} and {\it Living Cat} by orthogonal subspaces $M({\cal H})$ and $N({\cal H})$, respectively, where $M$ and $N$ are orthogonal projections, we can obviously construct a vector $|Cat\rangle$ that represents {\it Cat} with respect to {\it Dead Cat} and {\it Living Cat}, i.e. such that $\langle Cat|M|Cat\rangle=495,000/(495,000+29,400)=0.9439$ and $\langle Cat|N|Cat\rangle=29,400/(495,000+29,400)=0.0561$. Surprisingly, {\it Dead Cat} appears significantly more often than {\it Living Cat}. A likely explanation for this is that, more or less like newspapers, webpages are more often about events and situations that attract the attention of the reader, which is one of the reasons why the World Wide Web presents a specific human worldview. However, this is irrelevant to the explanatory framework that we put forward in this article, because the weights we find using Google do reveil the `meaning structure contained in the World Wide Web' with respect to {\it Cat}, {\it Dead Cat} and {\it Living Cat}, regardless of the nature of this meaning structure. After all, humans with minds more prone to be attracted by the remarkable may perhaps also opt for {\it Dead Cat} more often than for {\it Living Cat} if asked to make a choice..

We can therefore say that material entities that cannot communicate with human minds -- or, more generally, that cannot communicate with interfaces using semiotic signs, even when they are macroscopic in size, such as rocks, tables, chairs, etc\ldots -- are in fact entities of the microscopic realm. They consist of quantum entities, more specifically baryons, owing their stability -- and so their ability to become macroscopic in size -- from the Pauli exclusion principle \cite{dysonlenard1967,lenarddyson1968,lieb1976,lieb1979}. This exclusion principle also governs the organization of their being with respect to `place', i.e. it is impossible for two of them to occupy the same place in space. Classical physics `describes how such huge baryonic structures behave in space characterizing their states'. In principle, if we take into account the view on the micro and macro realms that we put forward, it is not surprising that a physics which focuses exclusively on the description of this Pauli exclusion principle mediated way of being for material entities is very different from quantum physics describing the dynamics of the quantum entities themselves and how they interact with their interfaces. To make this clear, let us consider the equivalent of this situation in the macroscopic realm. When someone dives into a pool, their head and body will move along a parabolic path, which is modeled quite well by classical mechanics describing the motion of a body in the gravitation field of the earth. Once the swimmer is in the water, classical fluid dynamics takes over and again provides a very good model for how the head and body move through the water. Of course, this classical dynamics will interfere with a quantum-like dynamics when during this activity the swimmer communicates with friends sitting beside the pool. The movement of the swimmer's head and body will be affected  by the exchange of meaning between the swimmer's mind and those of their friends. We can also imagine a voice coming from a loudspeaker for the swimmer to be heard, which would most probably influence the motion of the swimmer's head and body. The classical dynamics description of the swimmer's motion concerns most of the mass of the swimmer's head, and, for example, also the friction with the water, while for the quantum dynamics modeling the conversation between the different minds is the defining aspect, and the mass does not play an important role. The example of the swimmer proves that there is no contradiction for a reality where both types of dynamics take place jointly, influencing one another in a complicated contextual way. Taking into account our explanatory framework, we can say that this is what might also happen in the microscopic realm. Material entities, as oversized quantum memory structures, interact with each other in a realm most of all governed by gravitation, so that classical mechanics is a good model for their overall dynamics. They also interact with each other exchanging quantum particles continuously, and this interaction equally influences their behavior. But both realms co-exist without leading to contradictions.

This explains why it is necessary to realize very specific laboratory conditions for large-sized material entities of ordinary matter to reveal their microscopic quantum nature. One of the most spectacular examples is the realization of a Bose Einstein condensate at extremely low temperatures. A dilute gas of weakly interacting bosons confined in an external potential is cooled down to a temperature very near to absolute zero. Under such conditions a large fraction of the molecules of the gas are in the lowest quantum state of the external potential, and the wave functions of the different molecules overlap each other, and make the quantum effects which follow for the absolute identity of the molecules of the gas apparent on a macroscopic scale \cite{andersonenshermatthewswiemancornell1995,davismewesandrewsdrutendurfeekurnketterle1995}. Hence, this is an example of `microscopic quantum effects' becoming apparent on a macroscopic scale, not to be confused with `macroscopic quantum effects', as in our analysis of Schr\"odinger's Cat paradox. However, this also shows, following our basic hypothesis, that the conceptual nature of the gas consisting of bosons is its fundamental nature, which comes into the open only at very low temperatures. At higher temperatures, the  molecules are taking part in continuous interactions with other quantum particles, such that, looked at it from our explanatory perspective, their conceptual structure is very complex, and the aspect `identity' plays a minor role. Molecules behave almost as individuals -- in the `complicated and relative way' like we analyzed in section \ref{identityindividuality}, which is also the reason that Maxwell Boltzman statistical modeling gives good results for gases at higher temperatures.  

To end this section, we will pay attention to a type of quantum experiment where the remaining presence of the conceptual nature of a quantum particle, also in elaborate and complex situations, is put into evidence in a very convincing way, and this is the so-called quantum eraser experiment \cite{scullydruhl1982,yoonhoyukulikshihscully2000,walbornterracunhapaduamonken2002}. Two atoms $A$ and $B$ are excited by a laser pulse (see Figure 9).
\begin{figure}[H]
\centerline {\includegraphics[width=11cm]{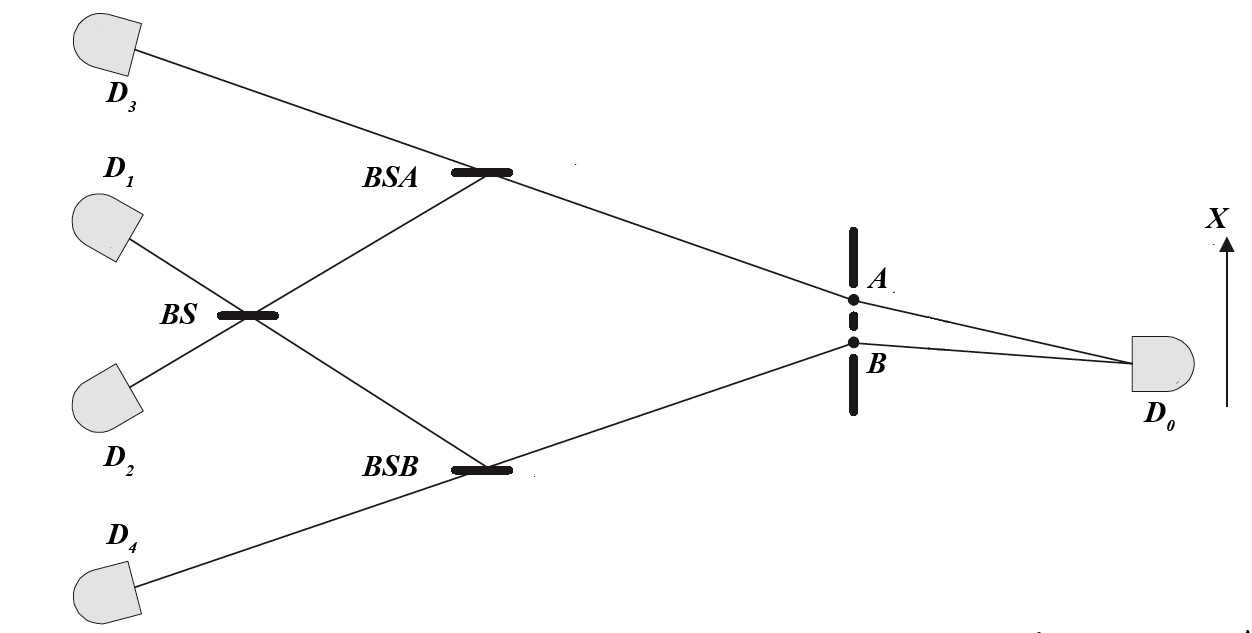}}
\caption{A proposed quantum eraser experiment. A pair of entangled photons is emitted from either atom $A$ or atom $B$ by atomic cascade decay. Clicks at $D_3$ or $D_4$ provide which-path information and hence lead to no interference at $D_0$. Clicks at $D_1$ or $D_2$ erase the
which-path information and lead to interference at $D_0$.}
\end{figure}
\noindent
A pair of entangled photons, photon 1 and photon 2, is then emitted
from either atom $A$ or atom $B$ by atomic cascade
decay. Photon 1, propagating to the right, is registered by a photon counting detector $D_0$, which can be scanned by a step motor along its $x$-axis for the observation of interference fringes. Photon 2, propagating to the left, is injected into a beamsplitter. If the pair is generated in
atom $A$, photon 2 will follow the $A$ path meeting $BSA$ with 50\% chance of being reflected or transmitted. If the pair is generated in atom $B$, photon 2 will follow the $B$ path meeting $BSB$ with 50\% chance of being reflected or transmitted. Under the 50\% chance of being transmitted by either $BSA$ or $BSB$, photon 2 is detected by either
detector $D_3$ or $D_4$. The registration of $D_3$ or $D_4$ provides which-path information (path $A$ or path $B$) of photon 2 and in turn provides which-path information of photon 1 because of the entanglement nature of the two-photon state of atomic cascade decay. Given a reflection at either
$BSA$ or $BSB$, photon 2 will continue to follow its $A$ path or $B$ path to meet another 50-50 beamsplitter $BS$ and then be detected by either detector $D_1$ or $D_2$, which are placed at the output ports of the beamsplitter $BS$. The
triggering of detectors $D_1$ or $D_2$ erases the which-path information, so that either the absence of the interference or the restoration of the interference can be arranged via an appropriately contrived photon correlation study. The basic idea for a quantum eraser experiment was proposed in \cite{scullydruhl1982}, and meanwhile real experiments have been performed confirming in a very convincing way the predictions of quantum mechanics \cite{yoonhoyukulikshihscully2000,walbornterracunhapaduamonken2002}.

Looking for an analogous situation within the realm of human concepts, we consider again the combination of concepts {\it The Animal eats the Food}, but now substituting {\it Food} by {\it Fruits or Vegetables}, so that the combination of concepts we consider is {\it The Animal eats Fruits or Vegetables}. Figure 10 is a schematic representation of the example we will make explicit now. In section \ref{entanglement}, we studied the entanglement of {\it Animal} and {\it Food} within the conceptual combination {\it The Animal eats the Food}, and hence this time we consider the entanglement between {\it Animal} and {\it Fruits or Vegetables} within the conceptual combination {\it The Animal eats Fruits or Vegetables}. In section \ref{interferencesuperposition}, we studied the interference of {\it Fruits} and {\it Vegetables}, establishing how {\it Fruits or Vegetables} is modeled by the superposition of {\it Fruits} and {\it Vegetables}. We will now consider how the combinations of concepts {\it The Animal is a Fructivore}, {\it The Animal is a Herbovire} and {\it The Animals is a Fructivore or a Herbivore} can be combined with the originally considered combination of concepts {\it The Animal eats Fruits or Vegetables}. This gives rise to three possibilities. The first two, {\it The Animal which is a Fructivore eats Fruits or Vegetables} and {\it The Animal which is a Herbivore eats Fruits or Vegetables}, will destroy the interference between {\it Fruits} and {\it Vegetables} modeled in section \ref{interferencesuperposition}.  
\begin{figure}[H]
\centerline {\includegraphics[width=11cm]{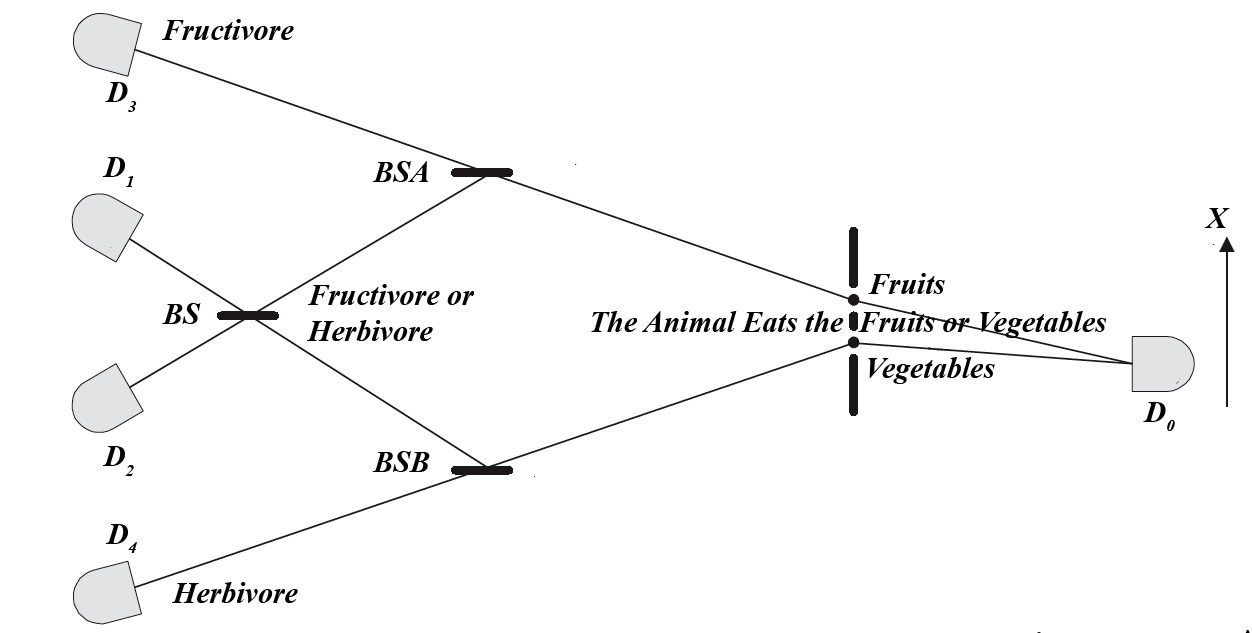}}
\caption{A schematic representation of the `human concept combination' equivalent of the quantum eraser. The entangled state is the combination of concepts {\it The Animal eats the Fruits or the Vegetables}. If {\it Animal} collapses to {\it Fructivore} (or {\it Herbivore}, respectively), {\it Fruits or Vegetables} collapses to {\it Fruits} (or {\it Vegetables}, respectively) and the interference disappears. However, if we delete the specification of {\it Animal}, so that the animal is {\it Fructivore or Herbivore}, the interference appears again.}
\end{figure}
\noindent
Indeed, if measurements  analogous to the ones carried out in \cite{hampton1988} are performed, where the subjects are asked to estimate the typicality weights of the exemplars in Table 1 for a random choice between one of the two combinations of concepts {\it The Animal which is a Fructivore eats Fruits or Vegetables} and {\it The Animal which is a Herbivore eats Fruits or Vegetables}, the outcomes will be comparable with the ones measured if subjects are asked to respond to a random choice of the combinations of the concepts {\it Fruits} and {\it Vegetables}. This means that a situation without interference, like the one graphically represented in Figure 6, will result. This situation is analogous to that in which detector $D_3$ or $D_4$ is triggered for the entangled photons, which indeed corresponds to a situation without interference. The third possibility is that subjects estimate typicality weights of the exemplars in Table 1 for the conceptual combination {\it The Animal which is a Fructivore or a Herbivore eats Fruits or the Vegetables}. Since the first part of sentence {\it The Animal which is a Fructivore or a Herbivore eats\ldots} does not imply anything for the second part of sentence {\it \dots Fruits or Vegetables}, we are confident that the interference modeled in section \ref{interferencesuperposition} is restored by these `human concept eraser experiments'.

This analysis of the quantum eraser serves to point out two aspects that are relevant to our explanatory framework. The first is that the mechanism of `erasing' works `because the conceptual combination remains a whole connected through meaning' all this time. Even if new pieces of combinations of concepts are added, `the field of meaning' immediately makes them flow into a whole. We believe that coherence must be interpreted in this way for the case of quantum particles. Of course, we can see that even in a seemingly innocent operation such as producing a figure like Figure 9, there is a deep-lying mistake of `not taking into account the nature of wholeness of the field of coherence' that has entered the entire approach, which is why the effect has been called `non-locality'. The second is that, should our basic hypothesis prove to be correct, namely that quantum particles are conceptual entities in the micro realm, the experiments of the quantum eraser type show that the conceptual content of these quantum particles is manifestly present also in complex experimental situations that cover a macroscopic scale.  

\subsection{Questions and Ideas for Future Investigation}

i) There is a strong and fundamental evolutionary aspect to our explanatory framework. Indeed, if quantum particles are conceptual entities with respect to interfaces which are entities of ordinary matter, this means that such interfaces of ordinary matter have co-evolved with, rather than evolved out of, such quantum particles. This co-evolution view could be the basis of a quite distinct way of looking at the evolution of the whole universe. For instance, dark matter might well be `that part of matter that did not take part in this co-evolution', and hence be the equivalent in the microscopic realm of what non-memory structures are in the macroscopic realm, which would explain its abundance as compared to ordinary matter. It would also mean that types of evolution in which also conceptual structures play a fundamental role, such as cultural evolution, would go much further back in time, having started already on the level of ordinary matter.

ii) Gravitation may well play a special role, also within the explanatory framework we put forward. The masses of quantum particles are not simple and entire multiples of a unit mass, which seems to indicate that mass is less directly connected to the notion of identity of concepts. This could be a consequence of gravity working on a very different scale of magnitude than the other forces. So possibly quantization, if connected to concept-like nature, has to be approached in a different way with respect to gravity.

iii) Although `meaning has the tendency to connect in a wholeness way all concentrations of meaning', there do exist separated and isolated concentrations of meaning in the realm of human concepts and their memory structures, for example two non communicating human minds. It will not be possible to model such isolated and separated concentrations of meaning in the vector space structure which models connected fields of meaning well. Could a more general mathematical structure, as the ones studied in quantum axiomatics, shed light on these situations \cite{aerts1999,aerts1994}?

iv) Quantum particles all show mass or energy when mass or energy is measured. Also human concepts make use of mass or energy carriers when they are exchanged between memory structures. We found an estimate of the energy involved in the pronunciation of a syllable to be $2\times10^{-5}$ Joules. Since one electron volt equals $1.602\times10^{-19}$ Joules, this means that a syllable carries an energy of around 124 TeV, where 1 TeV is $10^{12}$ electron volt. The rest mass-energy of an electron is $5.11$ KeV, where 1 KeV is $10^3$ electron volt. A muon has rest mass-energy $106$ MeV, where 1 MeV equals $10^6$ electron volt, and a tauon $1.78$ GeV, and 1 GeV equals $10^9$ electron volt. An up quark has rest mass-energy $1.9$ MeV, a down quark $4.4$ MeV, a strange quark $87$ MeV, a charm quark $1.32$ GeV, a bottom quark $4.24$ GeV, and a top quark $172.7$ GeV. This means that a syllable has a mass-energy which is more or less 1,000 times heavier than the heaviest elementary particle, the top quark. In turn, this top quark is about 10,000 times heavier than the lightest quark, the up quark, and 1,000 times heavier than the muon.

v) The generation problem is one of the mysterious and elusive situations of particle physics. Could the generations of the elementary particles, electron, muon, tauon, and their corresponding neutrinos and the different generations of quarks correspond to different energetic realizations of the conceptual structure of the quantum particles? It is true that human concepts have different mass-energetic realizations as well: a word can appear in sound-energetic form, but also in electromagnetic form when transported electronically or in writing, or in its primitive form used by our ancestors, carved into stone. All forms have different mass-energies, but, since they represent the same concepts, they have the same properties. Could quark confinement be the quantum particle equivalent of syllable and letter confinement for human concepts? Each human concept, when realized in a mass-energetic way, consists of syllables, and when written consists of even smaller units, namely letters. But there are no concepts corresponding to syllables or to letters because syllables and letters do not participate in the dynamics of meaning, hence they are confined. If the confinement of quarks would be due to quarks being building block of baryons, but not participating in the quantum coherence field, quarks would then be a consequence of the co-evolution between quantum particles as conceptual entities and ordinary matter as their memory structures, like syllables and letters are consequences of the co-evolution of human concepts and their memory structures.

\end{document}